\documentclass[10pt,journal,compsoc]{IEEEtran}
\pdfoutput=1
\ifCLASSOPTIONcompsoc
    \usepackage[nocompress]{cite}
\else
  \usepackage{cite}
\fi
\ifCLASSINFOpdf
\else
\fi
\usepackage{cite}
\usepackage{amsmath,amssymb,amsfonts}
\usepackage{amsthm,amsmath,amssymb}
\usepackage{dutchcal}
\usepackage{algorithmic}
\usepackage{algorithm}
\usepackage{graphicx}
\usepackage{textcomp}
\usepackage{xcolor}
\usepackage{diagbox} 
\usepackage{stfloats}
\usepackage[numbers,sort&compress]{natbib}
\usepackage{epstopdf}
\usepackage{multirow}
\usepackage{xcolor}
\usepackage{colortbl,booktabs}
\usepackage{subfigure}
\usepackage{caption}
\usepackage{extpfeil,extarrows}

\newtheorem{theorem}{Theorem}  [section]
\newtheorem{lemma}{Lemma}  [section]
\newtheorem{definition}{Definition}  [section]
\newtheorem{assumption}{Assumption}  [section]
\begin{document}
\title{Micro Analysis of Natural Forking in Blockchain Based on Large Deviation Theory}

\author{Hongwei~Shi,
        Shengling~Wang,~\IEEEmembership{Senior Member,~IEEE,}
        Qin~Hu,
        and~Xiuzhen~Cheng,~\IEEEmembership{Fellow,~IEEE}
\IEEEcompsocitemizethanks{
\IEEEcompsocthanksitem Hongwei Shi is with the School of Artificial Intelligence, Beijing Normal University, Beijing, China. E-mail: hongweishi@mail.bnu.edu.cn.\protect\\

\IEEEcompsocthanksitem Shengling Wang (Corresponding author) is with the School of Artificial Intelligence and the Faculty of Education, Beijing Normal University, Beijing, China. E-mail: wangshengling@bnu.edu.cn.\protect\\

\IEEEcompsocthanksitem Qin Hu is with the Department of Computer and Information Science, Indiana University - Purdue University Indianapolis, IN, USA. E-mail: qinhu@iu.edu.\protect\\

\IEEEcompsocthanksitem Xiuzhen Cheng is with the School of Computer Science and Technology, Shandong University (SDU), Shandong, China. E-mail: xzcheng@sdu.edu.cn.\protect\\

}
\thanks{}}

\markboth{Journal of \LaTeX\ Class Files,~Vol.~14, No.~8, August~2015}%
{Shell \MakeLowercase{\textit{et al.}}: Bare Demo of IEEEtran.cls for Computer Society Journals}
\IEEEtitleabstractindextext{%
\begin{abstract}
Natural forking in blockchain refers to a phenomenon that there are a set of blocks at one block height at the same time, implying that various nodes have different perspectives of the main chain. Natural forking might give rise to multiple adverse impacts on blockchain, jeopardizing the performance and security of the system consequently. However, the ongoing literature in analyzing natural forking is mainly from the macro point of view, which is not sufficient to incisively understand this phenomenon. In this paper, we fill this gap through leveraging the large deviation theory to conduct a microscopic study of natural forking, which resorts to investigating the instantaneous difference between block generation and dissemination in blockchain. Our work is derived  comprehensively and complementarily via a three-step process, where both the natural forking probability and its decay rate are presented. Through solid theoretical derivation and extensive numerical simulations, we find 1) the probability of the mismatch between block generation and dissemination exceeding a given threshold dwindles exponentially with the increase of natural forking robustness related parameter or the difference between the block dissemination rate and block creation rate; 2) the natural forking robustness related parameter may emphasize a more dominant effect on accelerating the abortion of natural forking in some cases; 3) when the self-correlated block generation rate is depicted as the stationary autoregressive process with a scaling parameter, it is found that setting a lower scaling parameter may speed up the failure of natural forking. These findings  are valuable since they offer a fresh theoretical basis to engineer optimal countermeasures for thwarting natural forking and thereby enlivening the blockchain network.
\end{abstract}
}

\maketitle

\IEEEpeerreviewmaketitle

\IEEEraisesectionheading{\section{Introduction}\label{sec:introduction}}
\IEEEPARstart{B}{lockchain} has recently obtained popular fascination for its use in digital cryptocurrency systems \cite{coexistence, bitcoin}, Internet of Things \cite{queuing}, and many other substantial fields \cite{lifesaving, public}. Essentially, blockchain is a ``state machine replication" \cite{quest}, which can facilitate confidence in a decentralized zero-trust context via {\it consensus} \cite{lya}. As the cornerstone of blockchain security, the consensus mechanism is called for reaching consistency among nodes on each state of the public ledger, indicating that all the nodes agree upon which block is a valid one. Any divergence in this process indicates the break of consensus, triggering the collapse of distributed trust. To select a valid block, blockchain resorts to the ``longest chain rule", which means that only if a block is appended in the longest chain (main chain), can this block be deemed as a valid one.

However, there may arise ambiguity among nodes over which the longest chain is due to the ineffective synchronization network. Notably, the underlying overlay network of blockchain is the P2P network, that may be annoyed by inefficient transmission and thereby imposing different views on the main chain among nodes, resulting in {\it natural forking}. Technically speaking, natural forking refers to a phenomenon that there are a set of blocks $\mathcal{B}_h$ at the same block height $h$ at a time, i.e., $|\mathcal{B}_h|>1$, implying that various nodes have different perspectives of the main chain.

Natural forking might give rise to multiple adverse impacts on blockchain, jeopardizing the performance and security of the system consequently. Firstly, it may cause low resource utilization. Owing to natural forking, nodes may perceive different recognition of the main chain and bury themselves in consuming computing power on the chain that is destined to be invalid, leading to a huge waste of computing power. Secondly, it provides a breeding ground for other mining attacks since the unexpected forks may disperse honest computing powers and thus being further manipulated to achieve malevolent intentions, such as double spending \cite{bitcoin} and selfish mining attack \cite{majority}. Hence, through exploiting natural forking, the blockchain network might execute in a resource-wasting and insecure manner.

As a result, it is pivotal for blockchain architects to have a good command of the reasons and consequences of natural forking, based on which, they can predict the occurrence of natural forking and thereby adopting countermeasures. Basically, the ongoing literature in depicting natural forking is mainly from the macro point of view \cite{p2p,2019-blockchain-2,2020-intention,2019ETH,xiao2020,TNSE}. That is, they carried out investigations based on the measurement of how a block is generated and disseminated systematically, and macroscopically derived the forking rate of the blockchain system.

Nonetheless, inspecting natural forking merely from the macro perspective is not sufficient to incisively understand this phenomenon. In general, the network transmission capability is beyond  the block generation capacity, enabling the network to transfer generated blocks timely. Even so, natural forking may still exist. This is because the {\it instantaneous difference} between block generation and dissemination will temporarily overturn the advantageous position of network transmission due to the randomness of the block generation and dissemination processes. And it is this instantaneous difference that incurs natural forking unexpectedly. Hence, there is a need to fill this gap by proposing a microscopic study of natural forking to analyze the short-term effects of the mismatch between block generation and dissemination. By doing so, we can realize more accurate observations of natural forking, which motivates our analysis in this work.

We face three challenges in studying natural forking microscopically. First of all, natural forking could be attributed to the fact that created blocks can not be propagated to other nodes timely. And such unsynchronized blocks might incur inconsistent recognition of the main chain among nodes, introducing natural forking. As such, we need to answer {\it how to dynamically and accurately quantify the mismatch between block generation and dissemination?} In addition, not all the mismatch between block generation and dissemination will arouse natural forking considering the {\it majority principle} in blockchain. In other words, blockchain is somewhat resistant to natural forking, and bifurcations are provoked only when such a mismatch exceeds a threshold. With this in mind, we need to analyze {\it whether a mismatch will induce natural forking or not?} Furthermore, given the dynamic characteristic of the natural forking process, we need to concentrate not only on the transitory forking rate, but also on its changing trend. Based on this, we need to respond to {\it how fast will the natural forking become severe?}

To address the first challenge, we translate the blockchain overlay network into a ``service system" which aims at propagating generated blocks to other nodes timely. As such, its input and output  can be regarded as the generated blocks (i.e., demand) and the disseminated blocks (i.e., supply). If the demand exceeds the supply, some nodes will remain unnoticed with the newly created blocks and the views of the longest chain from various nodes are inconsistent, making the blockchain network potentially threatened by natural forking and vice versa. Considering this, we construct a {\it queuing} model to depict the {\it supply and demand} problem in the service system and utilize the queue {\it backlog} to represent the mismatch between block generation and dissemination. To tackle the second challenge, a concept of {\it inconsistency-resistance degree} is introduced to denote the maximum number of blocks that have been created but are not disseminated without causing natural forking. Accordingly, only the mismatch is larger than the inconsistency-resistance degree, can it really incurs natural forking. Hence, this term benefits us from filtering out those null mismatches that do not produce natural forking. To solve the third challenge, we take advantage of the large deviation theory \cite{big} to conduct microscopical analysis. This enables us to scrutinize the probability of natural forking as well as its decay rate in a fine-grained way.

Our analysis is derived from simple to complex via  a three-step process which covers the single-source i.i.d. scheme, the single-source non-i.i.d. scheme, and the many-source non-i.i.d. scheme. Detailedly, in the first scheme, all the nodes behave as a whole (i.e., single-source) to produce newly mined blocks, and the number of generated blocks in each period is a sequence of independent and identically distributed (i.i.d.) random variables. As for the second scheme, all the nodes are also recognized as a single source yet its block generation rate within each interval is correlated to that of others (i.e., non-i.i.d.). Concerning the third scheme, every single node is regarded as one specific source (i.e., many-source) that inputs new blocks in a non-i.i.d. manner. As such, we can present a comprehensive and complementary study on the probability of natural forking and its decay rate. Conclusively, our contributions in this paper can be summarized as follows:
\begin{itemize}
   \item The analytical queuing model which takes block generation and dissemination as the input and output is studied from the micro point of view for the first time. Combined with the inconsistency-resistance degree, the probability of blockchain being threatened by natural forking is deduced in all three schemes.
   \item The decay rate of the probability that blockchain is threatened by natural forking is obtained, either explicitly (for the first scheme) or implicitly (for the second and third schemes) via solid theoretical derivation and extensive numerical simulations.
   \item We introduce two valuable concepts: the effective inconsistency-resistance degree and the effective network transmission rate, which respectively denote the minimum number of inconsistent blocks created yet not synchronized, and the minimum transmission rate of the overlay network, so as to guarantee the forking probability will not exceed a given threshold. This offers a fresh theoretical basis to engineer optimal countermeasures of natural forking.
   \item We find that appropriately setting the natural forking robustness related parameter or the difference between the block dissemination rate and block creation rate will reduce forking probability. Particularly, the natural forking robustness related parameter may emphasize a more dominant effect on accelerating the abortion of natural forking for the single-source i.i.d. scheme. This facilitates the blockchain designers to devise powerful defensive schemes for thwarting natural forking operatively.
   \item Without losing generosity, we characterize the self-correlated block generation rate as the stationary autoregressive process with a scaling parameter, where we find the rise of such scaling parameter may trigger the decline of the decay rate almost in a negative logarithmical way. Hence, setting a lower scaling parameter may speed up the failure of natural forking. This result boosts a deeper understanding of natural forking and helps to conceive powerful preventions against it.
 \end{itemize}

The rest of the paper is organized as follows. We first overview the related literature in Section \ref{sec:related work}, and interpret preliminaries in Section \ref{sec:pre}. Based on this, the single-source i.i.d. scheme, the single-source non-i.i.d. scheme, and the many-source non-i.i.d. scheme are respectively modeled and analyzed theoretically and experimentally in Sections \ref{sec:danyuan iid}, \ref{sec:danyuan non-iid} and \ref{sec:duoyuan}. Finally, we conclude our paper in Section \ref{sec:conclusion}.

\section{Related Work}\label{sec:related work}
The state-of-the-art literature for characterizing natural forking mainly focuses on macroscopically analyzing the probability of the following event: a conflict block $B$ is found while the previous block $A$ with the same block height is propagated in the network. As the first work on analyzing Bitcoin from the networking perspective and investigating how the information propagation time affects blockchain forks, Decher {\it et al.} in \cite{p2p} modeled the probability of blockchain forking through combining the probability of a block mined by the network and the ratio of the informed nodes of the previous block $A$. Based on this, they conjectured that the propagation delay is the primary cause of blockchain natural forks.

In \cite{2019-blockchain-2}, Seike {\it et al.} followed a similar analysis pattern as \cite{p2p} and theoretically calculated the forking rate even when the block generation rate is high. According to this, they derived a closed-form lower bound for the cumulative distribution function (CDF) of the time for extending the global block height. Besides, Liu {\it et al.} in \cite{2020-intention} inspected the unintentional forking in two cases more detailedly, which are forking caused by transmission delay and forking caused by transmission failure. After that, the performance under different resource utilization rates, block generation times were investigated. Additionally, Shahsavari {\it et al.} proposed the fork occurrence probability as a function of block propagation time and inter-block time from the blockchain networking perspective. Then, they extended the wave-based model so as to study the simultaneous propagation of concurrent blocks more specifically. Parallel to the above efforts, the impact of network delay of Ethereum on the number of forks was also modeled and quantified in \cite{2019ETH}, in which a simplified forking model was deduced under the assumption that all the nodes share the same hash power and equal communication capability. Taking the network connection heterogeneity into consideration, Xiao {\it et al.} in \cite{xiao2020} studied the forking rate when the nodes have different communication capabilities and various chances of winning the fork race. In light of this, they provided a concrete case study on how an adversary exploits blockchain forks. Stressed more on the miner's mining process, Chen {\it et al.} \cite{TNSE} explored the probability of forking and concluded that such possibility depends not only on the miner's hash rate, but also on the hash rates of other miners in the entire blockchain network coherently, according to which, an evolutionary game to scrutinize the competition dynamics of the miners in blockchain was derived.

Conclusively, all the above ongoing research considers natural forking merely from the macroscopic point of view, which is insufficient to understand this phenomenon deeply. Therefore, it is necessary to propose a micro study of natural forking to realize more accurate observations of it.

\section{Preliminaries}\label{sec:pre}
\begin{table*}
\caption{Comparison of three schemes in our paper.}\label{table}
\begin{center}
\begin{tabular}{|c|c|c|c| p{2cm}|}
\hline
\textbf{Scheme} & \textbf{Number of sources $\mathcal{N}$} & \textbf{Input sequence} & \textbf{Sampling space of input} \\ \hline
Single-source i.i.d. scheme & $\mathcal{N}=1$ & i.i.d. & discrete\\
\hline
Single-source non-i.i.d. scheme & $\mathcal{N}=1$ & non-i.i.d. & continuous   \\
\hline
Many-source non-i.i.d. scheme& $\mathcal{N}\ge2$ & non-i.i.d.& continuous \\
\hline
\end{tabular}
\end{center}
\end{table*}

In this section, we first recap the intrinsic reason of natural forking, paving the way for conducting queuing analysis in the following sections. Then, we introduce the progressive organization of our work in detail to reflect our analysis framework on the whole, which is advanced from simple to complex.

Essentially, blockchain utilizes the P2P network to transmit newly mined blocks, serving for reconciling the possible conflicting ledgers. Fundamentally, the risk of natural forking stems from the fact that there are blocks which have been created but are not propagated to other nodes timely. And those ``unserved" blocks might introduce different recognition of the longest chain among nodes, triggering natural forking consequently. Aware of this, we attribute the intrinsic reason of natural forking to the imbalance between the number of blocks required to be disseminated (i.e., the demand) and that could be broadcasted (i.e., the supply). If the demand is larger than the supply, it indicates some nodes are unaware of the newly mined blocks. Thus, the views of the longest chain from various nodes are different, making the blockchain network potentially threatened by natural forking; otherwise, on the occasion that the supply exceeds the demand, it implies the network transmission capacity is satisfactory to disseminate every newly mined block to other nodes in time, making no room of the disagreement on the main chain. Notably, the conflict occurring in the above supply and demand problem can be well captured by the queuing theory when taking block generation and dissemination as the input and output respectively. After that, we leverage the large deviation theory to scrutinize the probability of natural forking and its decay rate concurrently. These results are valuable since they provide a specific clue to suppress natural forking operatively.

In the following, we detailedly introduce our analysis process in this paper, which is carried out from simple to complex. Assume there are $N$ nodes in the blockchain system, with each generating blocks in a time-slotted manner. We partition the whole time span $T$ into multiple equalized periods with each's length as $\tau$. Hence, we have $t=\frac{T}{\tau}$ intervals. We begin by regarding all the nodes as a whole, which generates $c_i$ new blocks altogether into the system as a single source in the $i^{th}$ period  and $(c_i, i\in[1,t])$ is a sequence of independent and identically distributed (i.e., i.i.d.) random variables sampled from a discrete space. This is the simplest {\it single-source i.i.d. scheme} where the basic concepts and deviations will be introduced in Section \ref{sec:danyuan iid}. According to the inspections obtained from Section \ref{sec:danyuan iid}, we take a step further by investigating the {\it single-source non-i.i.d. scheme} in Section \ref{sec:danyuan non-iid}, where the block completion rates in different time intervals are correlated with each other and each of them is selected from a continuous space. This scheme is more practical in depicting the real blockchain system. In fact, the above two schemes regard the source of block generation as an individual one from the systematical perspective, which is insufficient to present a fine-grained analysis. Hence, we study the {\it many-source non-i.i.d. scheme} in Section \ref{sec:duoyuan} where every single node is considered as a specific source, injecting new blocks into the system in a non-i.i.d. manner. Particularly, the above three-step process proceeds complementarily and progressively since it explores the blockchain system from the single-source to many-source scenario, considers both i.i.d. and non-i.i.d. behaviors from discrete and continuous spaces. Comparison of the above three schemes in our paper is presented in Table. \ref{table}.

\textbf{Remark.} It is worth noting that in this paper, we do not consider the {\it delay attack} launched by a malicious adversary who might intentionally postpone some blocks so as to increase the probability of its block being appended into the main chain. However, our analysis based on the queuing theory and large deviation theory can be applied to such a scenario. This is because the essence of delay attack is to deliberately set the transmission capability of the target link to benefit the attacker. Thus, through adjusting the output strategy of the queuing model in our paper, our analytical model can also demonstrate the inconsistent phenomenon triggered by delay attack. As such, our work presents a general framework for investigating forks caused by network transmission, either honestly or maliciously.

\section{Single-Source I.I.D. Scheme}\label{sec:danyuan iid}
As stated above, we translate the underlying reason of natural forking to the mismatch between the number of blocks that should be and could be propagated. Once the transmission links in the network are capable of propagating messages timely, such a mismatch would be eliminated and thereby hampering natural forking. Aware of this, we are going to analyze the simplest case in this section, where all the nodes are regarded as a whole to inject newly mined blocks into the system, acting as the input side, while all the links transferring the blocks for synchronization are defined as the output side, to simulate the process of block generation and dissemination in blockchain. In fact, the mining process of each node follows Bernoulli distribution, which indicates that the block completion can be viewed as a Poisson process, so does the network transmission process \cite{corking, xiao2020, federated}. Simple this case is, it paves the way for further analyses conducted in the following sections.

Let $c_i^r$ be the number of blocks created by node $r$ in the $i^{th}$ period and  $c_i=\sum_{r=1}^N c_i^r$ be the total number of blocks mined by the system in this period. We assume that in each period, $c_i$ is independent and identically distributed. Considering that each new block is required to be broadcasted to all other nodes except its generator, there are $a_i=(N-1)\cdot c_i$ nodes demand to be ``served" by the network  and we notate $A_t=a_1+...+a_i+...+a_t$ as the number of cumulative blocks needed to be disseminated during the period $[0,T)$ with expectation $\lambda>0$. Subsequently, let $b_j$ be the number of blocks that all the links over the network are capable of transferring in the  $j^{th}$ period. Hence, $B_t=b_1+...+b_j+...+b_t$ expresses the summing ability of the network in transferring blocks during period $[0,T)$ with expectation $\mu>0$. Based on this, we define $Q_t$ as the difference in the number of blocks created and broadcasted at the end of the $t^{th}$ period, which is,
\begin{equation}\label{eq1}
\begin{aligned}
Q_t&=A_t-B_t\\
&=(a_1+...+a_t)-(b_1+...+b_t),
\end{aligned}
\end{equation}
with $Q_0=0$. Essentially, $Q_t$ reflects the difference between block creation and network propagation, and such an imbalance may bring in the divergence of the view on the main chain among the nodes that have not been synchronized to date. When $Q_t$ reaches the highest value, the blockchain system may suffer the severest confusion about which the longest chain is, thus is extremely threatened by natural forking. Therefore, to investigate the {\it threatened level} of natural forking in blockchain, we proceed to analyze the maximum value of $Q_t$, which is denoted as $Q=\sup_{t\ge0}Q_t$. Intuitively, $Q$ suggests the maximum number of unserved blocks of the system when $t\to\infty$, which we denote as the maximum number of {\it inconsistent} blocks in blockchain.

In fact, whether blockchain is veritably threatened by natural forking depends not solely on $Q$, but also on the security mechanism implemented in the system. For now, blockchain defaults to the {\it majority principle}, that is, as long as the majority of nodes (saying, half of the total nodes) receive the newest blocks timely and reach consensus on the longest chain, the system can escape from natural forking. In other words, blockchain is tolerable to bifurcation even if realizing the existence of inconsistent blocks. Hence, it is imperative to inspect the probability of $Q>\Omega$, denoted as $P(Q>\Omega)$, where $\Omega$ is described in the following definition:

\begin{definition}[Inconsistency-resistance degree] The inconsistency-resistance degree $\Omega>0$ is the maximum number of blocks created but not disseminated without causing natural forking of blockchain.
\end{definition}

In light of the above definition, $\Omega$ demonstrates the tolerable number of inconsistent blocks in the system when no natural forking occurs. In addition, it also presents the difficulty of launching natural forking. That is, the bigger $\Omega$ is, the more immune the system is to natural forking and vice versa. Hence, the probability $P(Q>\Omega)$ corresponds to the {\it threatened level} of the network by natural forking, which is a crux for us to analyze the robustness of blockchain.

Assume $\Omega=l\cdot q$ with $l \in \mathbb{N}$ and $q>0$. Based on Cramer's theorem in \cite{big}, if we take $l \to \infty$, $Q$ satisfies the large deviations principle with rate function $I(q)$, that is,
\begin{equation}\label{p}
\begin{aligned}
\lim_{l\to\infty}\frac{1}{l}\log P(Q>\Omega)=-I(q),
\end{aligned}
\end{equation}
in which
\begin{equation}\label{eq3}
\begin{aligned}
I(q)&=\mathop{\inf}_{t>0}\sup_{\theta\ge0}\{\theta \cdot q-t\cdot \Lambda(\theta)\}\\
&=\mathop{\inf}_{t>0}t\cdot\Lambda^*(\frac{q}{t}).
\end{aligned}
\end{equation}
In \eqref{eq3}, $\Lambda(\cdot)$ and $\Lambda^*(\cdot)$ represent the cumulant generating function of $Q_t$ and the convex conjugate of it respectively, which can be calculated by $\Lambda(\theta)=\mathop{\lim}_{t\to\infty}\frac{1}{t}\log \mathbf{E}[e^{\theta Q_t}]$ and $\Lambda^*(x)=\sup_{\theta\in\mathbb{R}}\{\theta x-\Lambda(\theta)\}$.

Since $Q_t=A_t-B_t$ in terms of \eqref{eq1} and the sequences $A_t$ and $B_t$ are independent, $\Lambda(\theta)$ can be derived as
\begin{equation}\label{eq4}
\begin{aligned}
\Lambda(\theta)&=\mathop{\lim}_{t\to\infty}\frac{1}{t}\log \mathbf{E}[e^{\theta (A_t-B_t)}]\\
&=\mathop{\lim}_{t\to\infty}\frac{1}{t}\log \mathbf{E}[e^{\theta A_t}]+\mathop{\lim}_{t\to\infty}\frac{1}{t}\log \mathbf{E}[e^{-\theta B_t}]\\
&=\Lambda_A(\theta)+\Lambda_B(-\theta),
\end{aligned}
\end{equation}
where $\Lambda_A(\theta)$ and $\Lambda_B(-\theta)$ respectively denote the cumulant generating function of $A_t$ and $B_t$. Based on this, standing for the convex conjugate of $\Lambda(\cdot)$, for $\forall x\in\mathbb{R}$, $\Lambda^*(x)$ can be derived as \cite{big}
\begin{equation}\label{convex}
\begin{aligned}
\Lambda^*(x)=\inf_{y\in\mathbb{R}}\{\Lambda_A^*(y)+\Lambda_B^*(y-x)\}.
\end{aligned}
\end{equation}

Hence, for obtaining $\Lambda^*(x)$ to determine $I(q)$ in \eqref{eq3}, we need to derive $\Lambda_A^*(\cdot)$ and $\Lambda_B^*(\cdot)$ respectively, which are summarized in the following lemma.

\begin{lemma}\label{lemma4.1}
$\forall x\ge 0$, we have
\begin{equation}\label{convex_A}
\begin{aligned}
\Lambda_A^*(x)=x\log(\frac{x}{\lambda})+\lambda-x,
\end{aligned}
\end{equation}

\begin{equation}\label{convex_B}
\begin{aligned}
\Lambda_B^*(x)=x\log(\frac{x}{\mu})+\mu-x.
\end{aligned}
\end{equation}
\end{lemma}

\begin{IEEEproof}
To prove \eqref{convex_A}, we are going to i) derive the explicit form of $\Lambda_A(\theta)$ and then ii) calculate $\Lambda_A^*(x)$ through $\Lambda_A^*(x)=\sup_{\theta\in\mathbb{R}}\{\theta x-\Lambda_A(\theta)\}$ accordingly.

i) To begin with, for the Poisson process $A_t$ with expectation $\lambda$, we have
\begin{equation}\label{proof1}
\begin{aligned}
\Lambda_A(\theta)&=\mathop{\lim}_{t\to\infty}\frac{1}{t}\log \mathbf{E}[e^{\theta A_t}]\\
&=\mathop{\lim}_{t\to\infty}\frac{1}{t}\log \mathbf{E}[e^{\theta \Sigma_{i=1}^t a_i}]\\
&=\log \mathbf{E}[e^{\theta a_1}]\\
&{\xlongequal{a_1\sim P(\lambda)}{}}\log \mathop{\Sigma}_{\kappa=0}^\infty (e^{\theta \kappa}\cdot \frac{\lambda^\kappa}{\kappa!}e^{-\lambda})\\ 
&=\log e^{\lambda (e^{\theta}-1)}\\
&=\lambda (e^{\theta}-1).
\end{aligned}
\end{equation}

ii) Subsequently, let $\phi(\theta)=\theta x- \lambda e^{\theta}+\lambda$, we then need to derive the maximum value of $\phi(\theta)$. Considering $\frac{\partial \phi(\theta)}{\partial{\theta}}=x-\lambda e^{\theta}$ and $\frac{\partial ^2 \phi(\theta)}{\partial{\theta}^2}=-\lambda e^{\theta}$, $\Lambda_A^*(x)$ gets its supremum as $\Lambda_A^*(x)=x\log(\frac{x}{\lambda})+\lambda-x$ when $\frac{\partial \phi(\theta)}{\partial{\theta}}=0$. Hence, we can conclude that \eqref{convex_A} holds. The proof procedure of $\Lambda_B^*(x)$ can be conducted similarly, thus we omit it to avoid redundancy.
\end{IEEEproof}

In light of Lemma \ref{lemma4.1}, $\Lambda^*(x)$ turns to
\begin{equation}
\begin{aligned}
\Lambda^*(x)&=\inf_{y\in\mathbb{R}} \varphi (y) = \inf_{y\in\mathbb{R}} [y\log(\frac{y}{\lambda})+\lambda-y\\
&+(y-x)\log\frac{(y-x)}{\mu}+\mu-(y-x)],
\end{aligned}
\end{equation}
where $\frac{\partial{\varphi(y)}}{\partial y}=\log \frac{y}{\lambda}+\log \frac{y-x}{\mu}$ and $\frac{\partial^2{\varphi(y)}}{\partial y^2}=\frac{1}{y}+\frac{1}{y-x}$ when $y>x>0$ is satisfied. Therefore, $\Lambda^*(x)$ can be derived as
\begin{equation}\label{derive}
\begin{aligned}
\Lambda^*(x)&=\lambda+\mu-\psi\\
&+\frac{1}{2}\psi\log \frac{x+\psi}{2\lambda}+\frac{x}{2}\log \frac{x+\psi}{2\lambda}\\
&+\frac{1}{2}\psi\log \frac{\psi-x}{2\mu}-\frac{x}{2}\log \frac{\psi-x}{2\mu}\\
&=\lambda+\mu-\psi+x\log\frac{x+\psi}{2\lambda},
\end{aligned}
\end{equation}
in which $\psi=\sqrt{x^2+4\lambda\mu}$.

Up to now, we have derived the explicit form of $\Lambda^*(x)$, according to which, $I(q)=\mathop{\inf}_{t>0}t\cdot\Lambda^*(\frac{q}{t})$ can be obtained. To be concrete, notate $\iota=t\cdot\Lambda^*(\frac{q}{t})$ through alternating $x$ with $\frac{q}{t}$ in \eqref{derive}. Thus, $\frac{\partial \iota}{\partial t}=\lambda+\mu-\sqrt{4\lambda\mu+\frac{q^2}{t^2}}$ and
$\frac{\partial^2 \iota}{\partial t^2}=\frac{q^2}{t^3\sqrt{4\lambda\mu+\frac{q^2}{t^2}}}>0$. Then we can get $I(q)$, i.e., the infimum of $\iota$, as
\begin{equation}\label{I(q)}
I(q)=q\log\frac{\mu}{\lambda}.
\end{equation}

Notably, we can state that the probability of the mismatch between block generation and dissemination exceeding a given threshold, i.e., $P(Q>\Omega)$, dwindles exponentially with $I(q)$ according to the large deviations principle. In this case, the rate function $I(q)$ shown in \eqref{I(q)} plays a critical role in representing the decay speed of the probability that natural forking actually happens. That is, if $I(q)$ is small, then $P(Q>\Omega)$ becomes large, indicating a higher probability of occurring natural forking. When this happens, it is necessary to take countermeasures for boycotting natural forking as soon as possible.

\begin{figure}[t]
\centering
\includegraphics[width=3in]{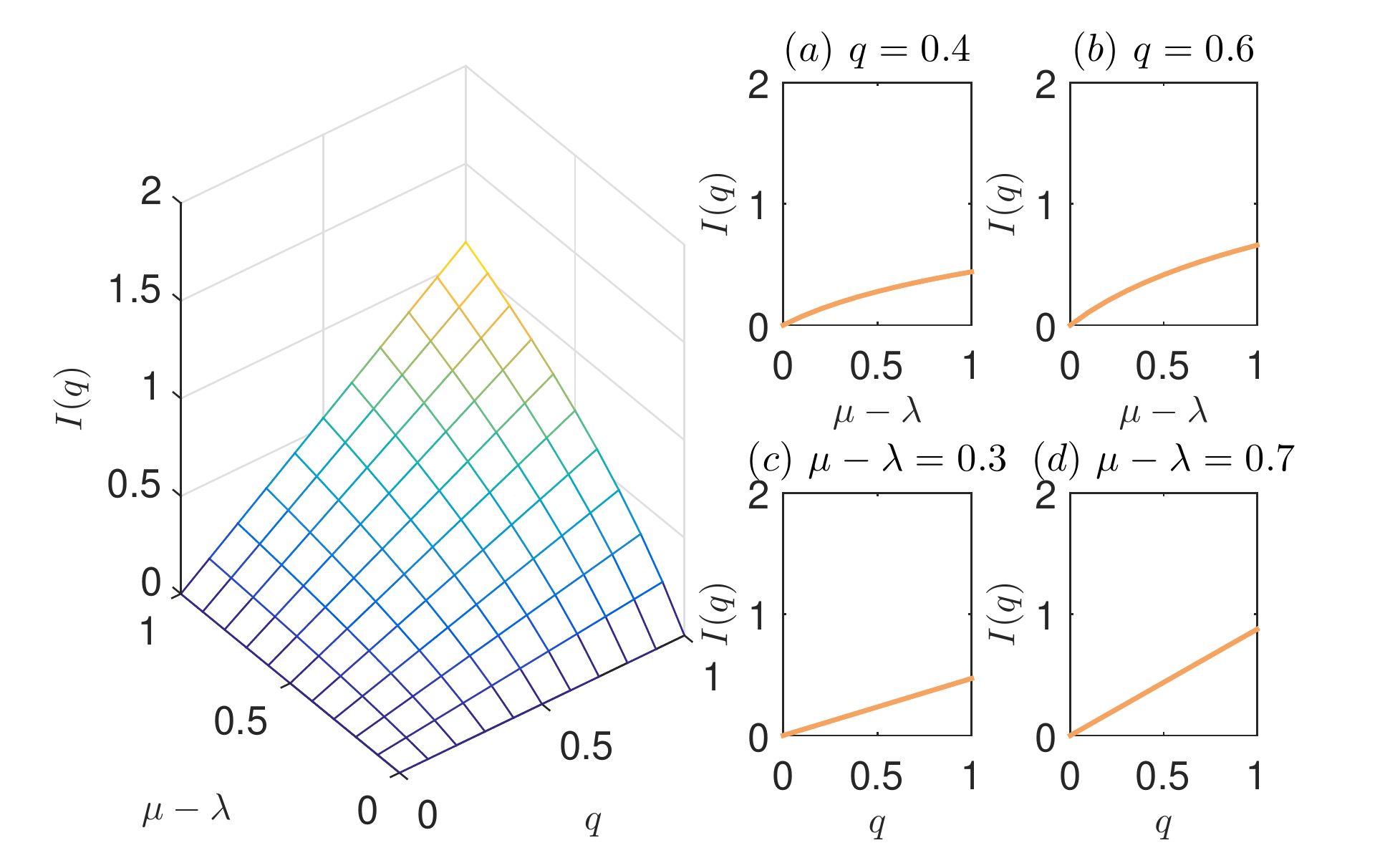}
\caption{The evolution of $I(q)$ in the single-source i.i.d. scheme.}
\label{fig:Iq}
\end{figure}

From \eqref{I(q)} and Fig. \ref{fig:Iq}, we can conclude that $I(q)$ rises when $q$ or the difference between $\mu$ and $\lambda$ increases, but differently in a linear or logarithmical way. This reflects that the former, i.e., the natural forking robustness related parameter $q$, stresses more on accelerating the abortion of natural forking, compared with the latter. In light of this, the blockchain system can be protected more effectively by setting $q$ appropriately instead of enhancing the difference between $\mu$ and $\lambda$. This insight edifies us to further investigate the best value of $\Omega$ considering its imperativeness in deciding $P(Q>\Omega)$, which is characterized in the following definition.

\begin{definition}[Effective inconsistency-resistance degree] The effective inconsistency-resistance degree $\Omega^*(\delta)$ is the minimum number of inconsistent blocks created but not disseminated in the network, to ensure the probability of natural forking not higher than a given threshold $\delta \in (0,1]$. That is,
$$\Omega^*(\delta)=\min\{\Omega:P(Q>\Omega)\le\delta\}.$$
\end{definition}

The $\Omega^*(\delta)$ defined above can be calculated detailedly by the following theorem.

\begin{theorem}\label{3.1}
The effective inconsistency-resistance degree $\Omega^*(\delta)$ of blockchain can be deduced by $\Omega^*(\delta)=-\frac{\log\delta}{\log\frac{\mu}{\lambda}}$, provided with $\delta\in(0,1]$ and $\mu>\lambda>0$.
\end{theorem}

Even if a larger $q$ can achieve better performance in resisting natural forking, it can not radically solve this problem. Tracing the root of natural forking occurrence, we hint that enhancing the network transmission ability to catch up with the block completion rate in real-time is substantial. Aware of this, we are facing the question that {\it what the network transmission rate should be so as to fight back natural forking?} To answer this, we present the following concept.
\begin{definition}[Effective network transmission rate]
The effective network transmission rate $\mu^*(\delta)$ denotes the minimum transmission rate of the blockchain network for ensuring the probability of natural forking no more than the given threshold $\delta\in(0,1]$. That is,
$$
\mu^*(\delta)=\min\{\mu:P(Q>\Omega)\le\delta\}.
$$
\end{definition}

Based on the definition of $\mu^*(\delta)$, we can derive the following theorem.

\begin{theorem}\label{4.2}
The effective network transmission rate $\mu^*$ can be calculated via $\mu^*(\delta)=\lambda e^{-\frac{\log \delta}{\Omega}}$, given $\lambda>0$ and $\delta\in(0,1]$.
\end{theorem}

Theorems \ref{3.1} and \ref{4.2} can be easily obtained, thus we omit their proofs for simplicity.

\section{Single-Source non-I.I.D. Scheme}\label{sec:danyuan non-iid}
In the above section, we derive the probability of natural forking $P(Q>\Omega)$ as well as its decay rate function $I(q)$. By doing so, we realize that natural bifurcation can be well thwarted by enhancing the robustness related parameter $q$. In addition, the analyses on the effective inconsistency-resistance degree $\Omega^*(\delta)$ and the effective network transmission rate $\mu^*(\delta)$ are also carried out so as to meet specific performance requirements on the probability of natural forking.

The traits of the above scheme are 1) all the nodes are regarded as a whole, i.e., a single source; 2) in each time interval $i\in[1,t]$, the number of created blocks $c_i$ is independent and identically distributed. However, these two assumptions are not perfectly apt to the real blockchain system. For the first assumption, if we unify all the nodes with different mining strategies as one single input, this one-size-fits-all method may introduce biases when inspecting $P(Q>\Omega)$. As for the second assumption, it may not hold when different mining strategies with multiple subjective purposes are employed during the mining process, making the block completion behavior of each node neither independently nor identically distributed. Representative mining strategies include the power adjusting withholding strategy \cite{PAW}, the pool-hopping strategy \cite{analysis}, and the action of mining gap \cite{gap}, just to name a few. With this in mind, a fine-grained analysis without being limited by these two assumptions should be conducted in order to fit in the practical blockchain world.

In this section, we relax the second hypothesis by regarding the number of generated blocks in each period as a self-related one. That is to say, the number of blocks mined in the $i^{th}$
interval is correlated to that of the $(i-1)^{th}$. By doing so, we can mimic the real blockchain system a step further. Based on this, we will lessen the first assumption  in the next section to approximate the real blockchain system finally.

Comparable to the analysis of single-source i.i.d. scheme, we suppose $\tilde{c_i}=\Sigma_{r=1}^N \tilde{c}_i^r$ be the total number of blocks generated by all blockchain nodes in the $i^{th}$ period in which  $\tilde{c}_i^r$ denotes the number of created blocks from node $r$. Different from the above scheme, we assume the total mining power of blockchain in each time interval is self-related. That is, $\tilde{c}_i$ evolves and adapts according to $\tilde{c}_{i-1}$ based on a map function $f$. Note that there is no specific restriction on how to depict $f$, as long as it meets the requirements of the corresponding scenarios, and the rigorous construction of $f$ is beyond the scope of our work. Hence, we do not focus on any specific function $f$ in this paper, while abstracting the essence of the self-adaptive nature by introducing $f$ as the {\it stationary autoregressive process of degree $1$} \cite{big} for simplicity. According to this, we can deduce the corresponding conclusions without losing generosity and it is also applicable to any other map function $f$.

With this in mind, the number of created blocks required to be broadcasted until the end of the $t^{th}$ period can be denoted as $\tilde{A_t}=\tilde{a_1}+...+\tilde{a_i}+...+\tilde{a_t}$ with $\tilde{a_i}=(N-1)\cdot\tilde{c_i}$. Evolving as the stationary autoregressive process \cite{big, auto1, auto2}, $\tilde{a_i}$ derives as
\begin{equation}\label{autoregressive}
\begin{cases}
\tilde{a_i}=\tilde\lambda+\chi_t, \\
\chi_t=\xi\cdot \chi_{t-1}+\sqrt{1-\xi^2}\cdot\tilde{\sigma}\cdot\varepsilon_t. \\
\end{cases}
\end{equation}

In \eqref{autoregressive}, $\tilde{\lambda}$ and $\tilde\sigma$ respectively represent the expectation and standard deviation of the number of blocks waiting to be propagated in the system. That is, $\mathbf{E} \tilde{a_i}=\tilde{\lambda}$ and $\mathbf{Var} \tilde{a_i}=\tilde\sigma^2$. Besides, $\chi_t$ expresses the bias of each $\tilde{a_i}$ in the $i^{th}$ period, which can be deduced according to $\chi_{t-1}$ previously, perturbed by a standard normal random variables $\varepsilon_t$. And $\xi$ is a scaling parameter with $\xi\in(-1,1)$.

Next, we define the number of blocks that the network could transfer as $\tilde b_j$ in the $j^{th}$ period with expectation $\tilde{\mu}>0$. Thus, the cumulative ability to transfer created blocks via links in the network in $[0,T)$ can be characterized by $\tilde{B_t}=\tilde b_1+...+\tilde b_j+...+\tilde b_t$. In this case, the difference in the number of blocks mined and disseminated until the end of the $t^{th}$ interval can be derived comparably as $\tilde{Q_t}$, that is,
\begin{equation}
\begin{aligned}
\tilde{Q_t}&=\tilde{A_t}-\tilde{B_t}\\
&=(\tilde{a_1}+...+\tilde{a_t})-(\tilde{b_1}+...+\tilde{b_t}),
\end{aligned}
\end{equation}
with $\tilde{Q_0}=0$. To investigate the threatened level by natural forking of blockchain, we turn to analyze the defined metric $\tilde{Q}=\sup_{t\ge0}\tilde{Q_t}$ in the following. In fact, the inconsistency-resistance degree $\tilde{\Omega}$ \footnote{The definition of $\tilde{\Omega}$ is the same as $\Omega$, but we change its notation for convenience of representation in this section to distinguish the analysis from the previous section, so does $\bar{\Omega}$ in the next section.} describes the largest tolerable number of blocks that have been created but can not be broadcasted without causing natural forking. If $\tilde{Q}>\tilde{\Omega}$, the blockchain system is vulnerable to natural forking, instructing us it is a must to take countermeasures for fighting back natural bifurcation.

Similarly, if we suppose $\tilde{\Omega}=r\cdot b$ with $r\to\infty$, inspired by Cramer's theorem, we can prove that $P(\tilde{Q}>\tilde{\Omega})$ satisfies
\begin{equation}\label{p2}
\begin{aligned}
\lim_{r\to\infty}\frac{1}{r}\log P(\tilde{Q}>\tilde{\Omega})=I(b),
\end{aligned}
\end{equation}
with the rate function $I(b)$ as
\begin{equation}\label{I(b)}
\begin{aligned}
I(b)=\mathop{\inf}_{t>0}t\cdot\tilde{\Lambda}^*(\frac{b}{t}).
\end{aligned}
\end{equation}

In \eqref{I(b)}, $\tilde{\Lambda}^*(\tilde{x})$ denotes the convex conjugate of the corresponding cumulant generating function of $\tilde{Q}_t$, i.e., $\tilde{\Lambda}(\theta)$, if we substitute $\frac{b}{t}$ as $\tilde{x}$. Hence, we have $\tilde{\Lambda}^*(\tilde{x})= \sup_{\theta\in\mathbb{R}}\{\theta \tilde{x}-\tilde{\Lambda}_A(\theta)-\tilde{\Lambda}_B(-\theta)\}=\mathop{\inf}_{\tilde{y}\in\mathbb{R}}\{\tilde{\Lambda}^*_A(\tilde{y})+\tilde{\Lambda}^*_B(\tilde{y}-\tilde{x})\}, \forall \tilde{x}\in\mathbb{R}$. Considering the definition of $\tilde{B_t}$ is analogous to that of $B_t$, thus,  $\tilde{\Lambda}^*_B(\cdot)$ is the same as $\Lambda_B^*(\cdot)$. However, $\tilde{A}_t$ is described differently as the stationary autoregressive process compared to $A_t$. Hence, we need to derive $\tilde{\Lambda}_A(\theta)$ and $\tilde{\Lambda}_A^*(\cdot)$ additionally, which is summarized in the following theorem.

\begin{theorem}
Given $\tilde{a}_i$ evolving as the stationary autoregressive process according to \eqref{autoregressive} in the $i^{th}$ time interval, the cumulant generating function of $\tilde{A}_t$ and its convex conjugate, i.e., $\tilde{\Lambda}_A(\theta)$ and $\tilde{\Lambda}^*_A(\tilde{x})$, can be calculated by
\begin{equation}\label{Lambda_A_wan}
\begin{aligned}
\tilde{\Lambda}_A(\theta)=\theta\tilde{\lambda}+\frac{\theta^2}{2}(\frac{1+\xi}{1-\xi}),
\end{aligned}
\end{equation}

\begin{equation}\label{Lambda*_A_wan}
\begin{aligned}
\tilde{\Lambda}^*_A(\tilde{x})=\frac{(\tilde{x}-\tilde{\lambda})^2(1-\xi)}{2(1+\xi)}.
\end{aligned}
\end{equation}
\end{theorem}

\begin{IEEEproof}
According to \eqref{autoregressive}, clearly we can get $\mathbf{E}{\tilde{A}_t}=\tilde{\lambda}t$ and $\mathbf{Cov}{(\tilde{a}_0,\tilde{a}_t)}=\xi^t\tilde\sigma^2$. Hence, we have $
\mathbf{Var}{\tilde{A}_t}=\tilde\sigma^2\mathop{\Sigma}_{1\le i,j\le t}\xi^{|i-j|}=\frac{\xi^2}{(1-\xi)^2}[t(1-\xi^2)-2\xi(1-\xi^t)]$. If we define $\tilde{\Lambda}_{A^t}(\theta)=\log \mathbf{E} [e^{\theta \tilde A_t}]$ and the limit of it as $\tilde \Lambda_A(\theta)=\lim_{t\to\infty}\frac{1}{t}\tilde\Lambda_{A^t}(\theta)$, then we have $\tilde{\Lambda}_{A^t}(\theta)=\theta\tilde\lambda t+\frac{1}{2}\tilde{\sigma}_t^2$. Hence, \eqref{Lambda_A_wan} can be derived through dividing by $t$ and taking limit $t\to\infty$.

As for \eqref{Lambda*_A_wan}, $\tilde{\Lambda}^*_A(\tilde{x})$ can be derived from $\tilde{\Lambda}^*_A(\tilde{x})=\sup_{\theta\in\mathbb{R}}\{\theta \tilde{x}-\tilde\Lambda_A(\theta)\}$. Denote $\rho(\theta)=\theta \tilde{x}-\theta \tilde{\lambda}-\frac{\theta^2}{2}(\frac{1+\xi}{1-\xi})$, we can derive $\frac{\partial\rho}{\partial\theta}=\tilde{x}-\tilde\lambda-\theta(\frac{1+\xi}{1-\xi})$ and $\frac{\partial^2\rho}{\partial\theta^2}=-\frac{1+\xi}{1-\xi}<0$. Therefore, $\rho(\theta)$ gets the maximum value as $\rho(\theta)^*=\frac{(\tilde{x}-\tilde\lambda)^2(1-\xi)}{2(1+\xi)}$ when $\theta=\frac{(\tilde{x}-\tilde\lambda)(1-\xi)}{1+\xi}$. Therefore, we finish the proof of this theorem.
\end{IEEEproof}

Recall that $\tilde{\Lambda}^*(\tilde{x})=\mathop{\inf}_{\tilde{y}\in\mathbb{R}}\{\tilde{\Lambda}^*_A(\tilde{y})+\tilde{\Lambda}^*_B(\tilde{y}-\tilde{x})\}, \forall \tilde{x}\in\mathbb{R}$, then $\tilde{\Lambda}^*(\tilde{x})$ turns to $\tilde{\Lambda}^*(\tilde{x})=\mathop{\inf}_{\tilde{y}\in\mathbb{R}} \Gamma(\tilde{y})$ where
\begin{equation}\label{gamma}
\begin{aligned}
\Gamma(\tilde{y})=\frac{(\tilde{y}-\tilde{\lambda})^2(1-\xi)}{2(1+\xi)}+(\tilde{y}-\tilde{x})\log\frac{\tilde{y}-\tilde{x}}{\tilde{\mu}}+\tilde{\mu}-(\tilde{y}-\tilde{x}).
\end{aligned}
\end{equation}
To obtain the minimum of $\Gamma(\tilde y)$, i.e., $\Gamma^*(\tilde{y})$, we derive partial derivatives as $\frac{\partial \Gamma(\tilde{y})}{\partial \tilde{y}}=\frac{1-\xi}{1+\xi}(\tilde{y}-\tilde\lambda)+\log\frac{\tilde{y}-\tilde{x}}{\tilde\mu}$ and $\frac{\partial^2 \Gamma}{\partial \tilde{y}^2}=\frac{1-\xi}{1+\xi}+\frac{1}{\tilde{y}-\tilde{x}}$. When $\tilde{y}>\tilde{x}>0$, we have $\frac{\partial^2 \Gamma}{\partial \tilde{y}^2}>0$, and $\Gamma^*(\tilde{y})$ can be acquired if
\begin{equation}\label{chaoyue}
\frac{1-\xi}{1+\xi}(\tilde{y}-\tilde\lambda)+\log\frac{\tilde{y}-\tilde{x}}{\tilde\mu}=0
\end{equation}
holds. However, \eqref{chaoyue} is a transcendental equation, we can not deduce its closed-form solution explicitly. Considering this, we conduct Taylor expansion on the logarithmic part of it. That is, we transform $\log\frac{\tilde{y}-\tilde{x}}{\tilde\mu}$ to $(\frac{\tilde{y}-\tilde{x}}{\tilde{\mu}}-1)-\frac{(\frac{\tilde{y}-\tilde{x}}{\tilde{\mu}}-1)^2}{2}$ for approximation. By doing so, we can easily find the solutions of $
\frac{1-\xi}{1+\xi}(\tilde{y}-\tilde\lambda)+(\frac{\tilde{y}-\tilde{x}}{\tilde{\mu}}-1)-\frac{(\frac{\tilde{y}-\tilde{x}}{\tilde{\mu}}-1)^2}{2}=0
$ as
\begin{equation}\label{solutions}
\begin{cases}
\tilde{y_1}=\frac{2\tilde{\mu}+2\xi\tilde{\mu}+\tilde{\mu}^2-\xi\tilde{\mu}^2+\tilde{x}+\xi \tilde{x}-\tilde{\mu}\pi}{1+\xi},\\
\tilde{y_2}=\frac{2\tilde{\mu}+2\xi\tilde{\mu}+\tilde{\mu}^2-\xi\tilde{\mu}^2+\tilde{x}+\xi \tilde{x}+\tilde{\mu}\pi}{1+\xi},
\end{cases}
\end{equation}
where $\pi=(1+2\xi+\xi^2-2\tilde{\lambda}+2\xi^2\tilde{\lambda}+4\tilde{\mu}-4\xi^2\tilde{\mu}+\tilde{\mu}^2-2\xi\tilde{\mu}^2+\xi^2\tilde{\mu}^2+2\tilde{x}-2\xi^2\tilde{x})^{1/2}$.
In the following, we need to substitute $\tilde{y}$ as $\tilde{y_1}$ and $\tilde{y_2}$ in \eqref{gamma} to get the minimum of $\Gamma(\tilde{y})$ so as to obtain $\tilde{\Lambda}^*(\tilde{x})$. After that, the rate function $I(b)$ can be derived through calculating the infimum according to \eqref{I(b)}.

\begin{figure}[tb]
\centering
\subfigure[$b=5,k=2$]{
\begin{minipage}[t]{0.45\linewidth}
\centering
\includegraphics[width=1.1\textwidth]{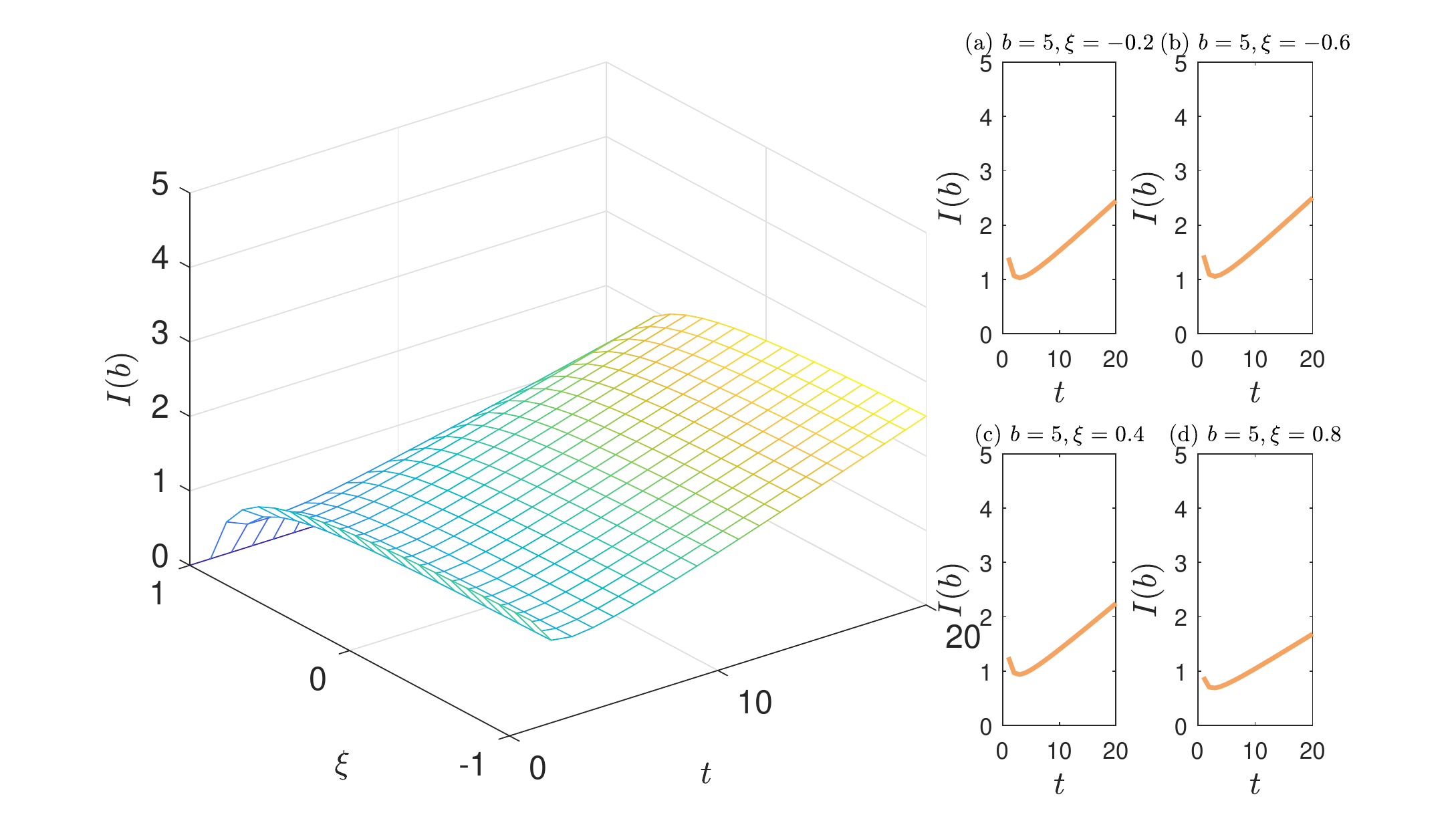}
\end{minipage}
}
\subfigure[$b=10,k=2$]{
\begin{minipage}[t]{0.45\linewidth}
\centering
\includegraphics[width=1.1\textwidth]{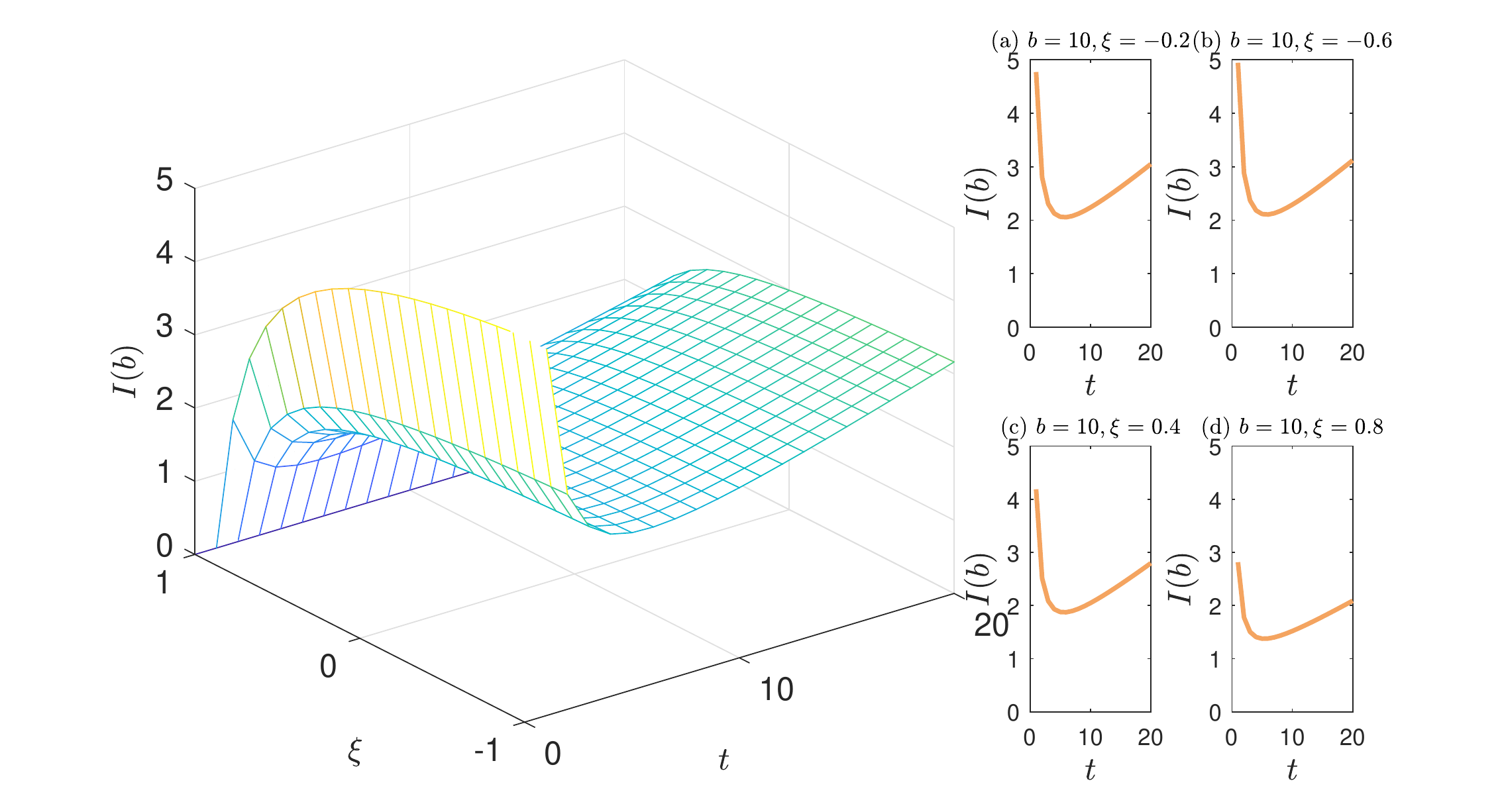}
\end{minipage}
}
\subfigure[$\xi=-0.2,k=2$]{
\begin{minipage}[t]{0.45\linewidth}
\centering
\includegraphics[width=1.1\textwidth]{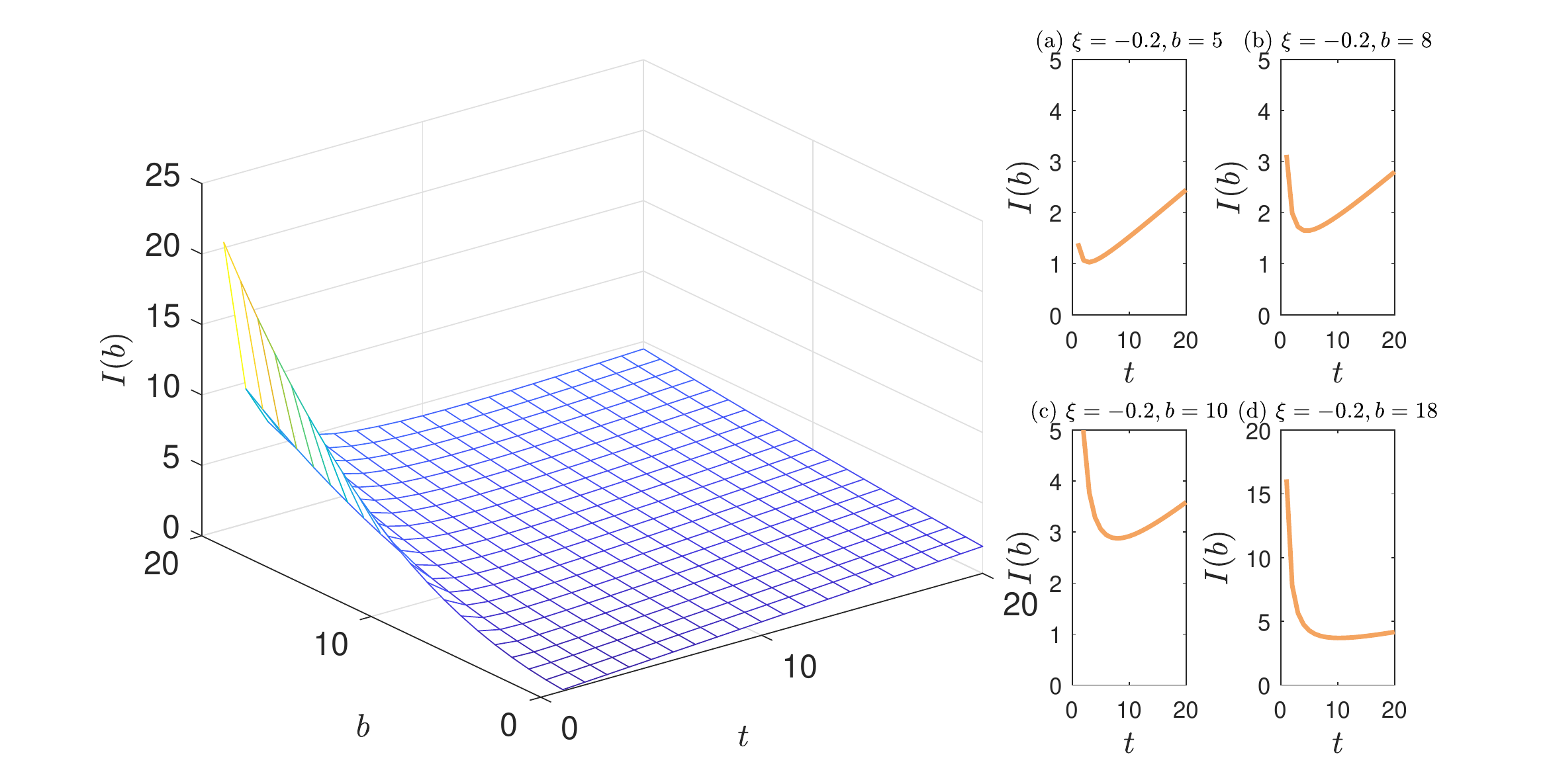}
\end{minipage}
}
\subfigure[$\xi=0.8,k=2$]{
\begin{minipage}[t]{0.45\linewidth}
\centering
\includegraphics[width=1.1\textwidth]{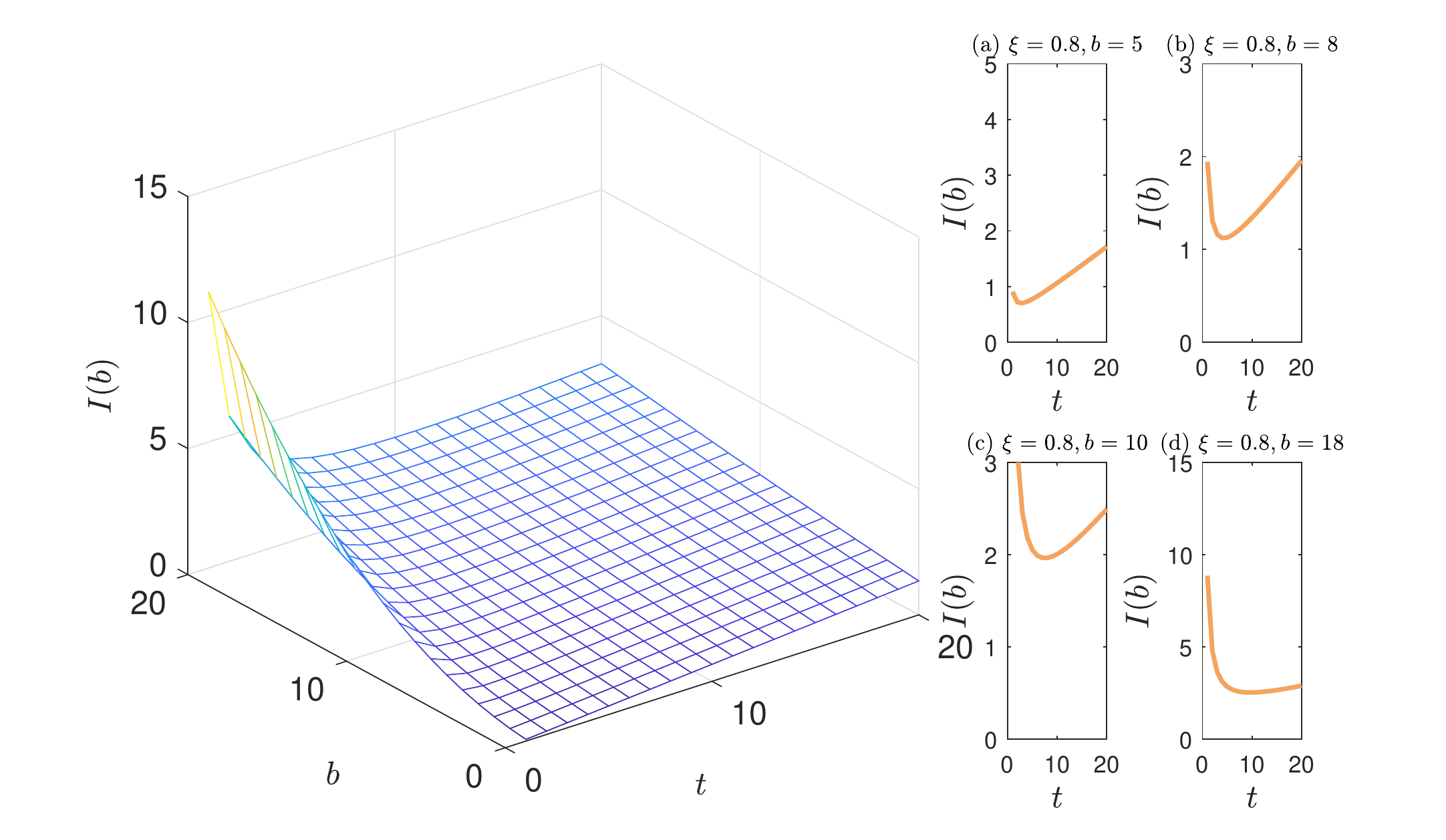}
\end{minipage}
}
\centering
\caption{The evolutions of $I(b)$ when $b=5,10$, $\xi=-0.2,0.8$ and fixed $k=2$.}
\label{fig:I(b)1}
\end{figure}

Nonetheless, the intricacy of \eqref{gamma} and \eqref{solutions} inevitably increase the computational complexity of the above calculation procedure and it is non-trivial to get the analytical solution of $I(b)$ consequently. In light of this, we conduct numerical simulations of $I(b)$ so as to obtain  insightful observations. Here we first interpret the parameter settings and experimental requirements for carrying out those simulations. To start with, we set $\tilde{y}=\tilde{y_1}$ in \eqref{gamma} to derive $I(b)$ since $\tilde{y}=\tilde{y_2}$ is incapable of producing the infimum of $\Gamma^(\tilde{y})$, that is, when $\tilde{y}=\tilde{y_2}$, $\tilde{y}\to+\infty$ leads to $I(b)\to-\infty$. Besides, to satisfy $\tilde{y}>\tilde{x}>0$, $\tilde{\lambda}>b$ should be met. In our simulations, we strictly comply with this requirement, and all the experiments are conducted under reasonable settings. Additionally, we carry out experiments when $t$ goes very large, i.e., $t=2000$. However, we only present the results of the first several rounds for getting a clear observation of $I(b)$. Finally, other parameter settings (i.e., $k\footnote{We denote the difference of the network transmission rate and that of the block completion as a fixed value $k$ for simplicity. The meaning of $h$ in the next section is the same.}=\tilde{\mu}-\tilde{\lambda}\in[0,2000]$, $b\in[0,500]$) have been tested extensively. And since they display very similar results, we omit them in this paper to reduce redundancy.

Our numerical simulations are performed in the following way. To begin with, considering $I(b)$ denotes the minimum of $t\cdot \tilde{\Lambda}^*(\frac{b}{t})$, we need to testify that such an infimum exists when $t$ varies, which is demonstrated in Fig.  \ref{fig:I(b)1}. More specifically, the subfigures (a) and (b) in  Fig.  \ref{fig:I(b)1} report the trends of $I(b)$ when $b=5$ and 10 with $k=2$ and $\xi\in(-1,1)$. Besides, subfigures (c) and (d) present the changes of $I(b)$ if $\xi=-0.2$ and 0.8 with $k=2$ and $b\in[0,20]$. Clearly, we can see that in all pictures, $I(b)$ descends first and increases subsequently as $t$ grows, implying there is an infimum lies in $I(b)$.

\begin{figure}[tpb]
\centering
\subfigure[$k=0.5$]{
\begin{minipage}[t]{0.45\linewidth}
\centering
\includegraphics[width=1.13\textwidth]{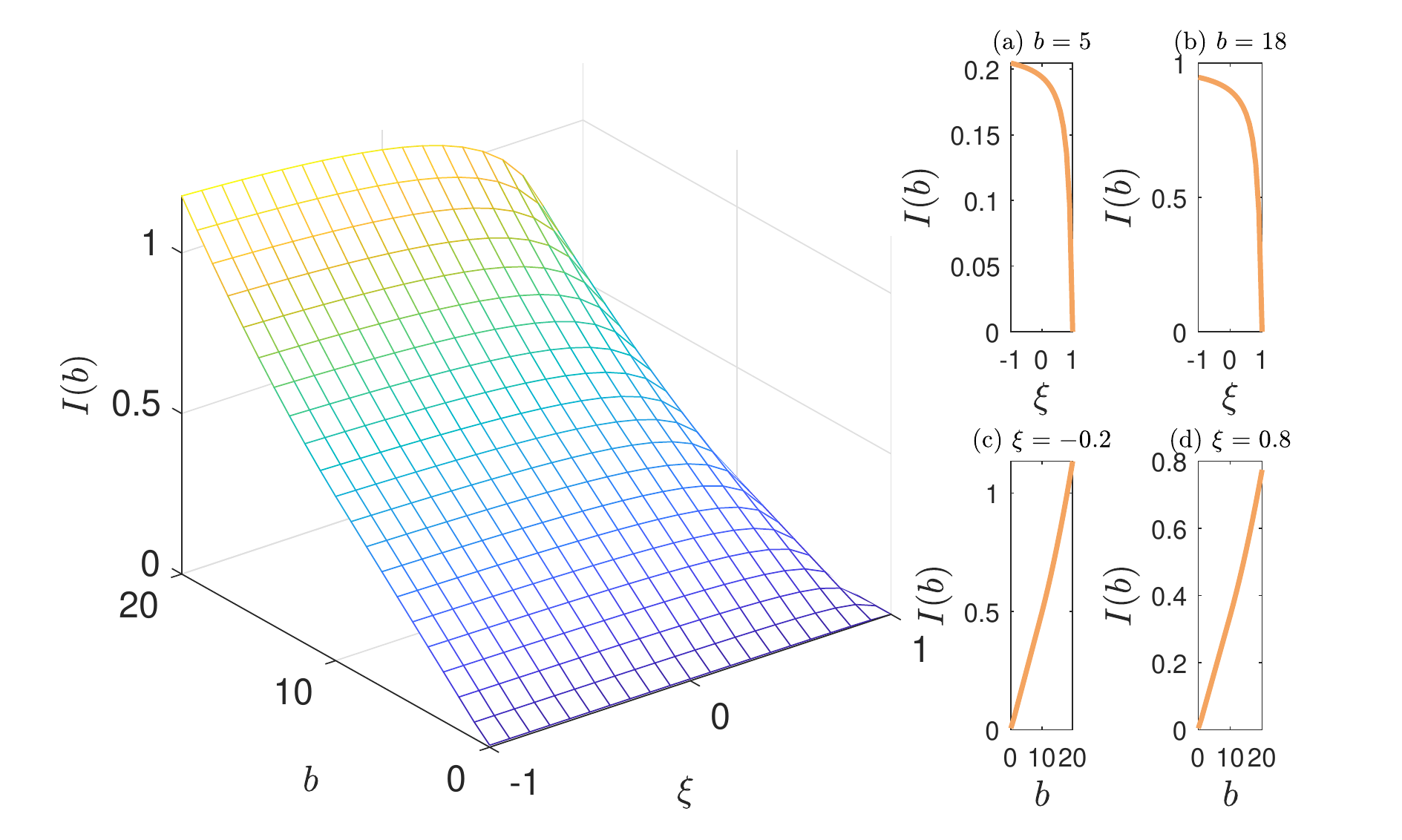}
\end{minipage}
}
\subfigure[$k=2$]{
\begin{minipage}[t]{0.45\linewidth}
\centering
\includegraphics[width=1.1\textwidth]{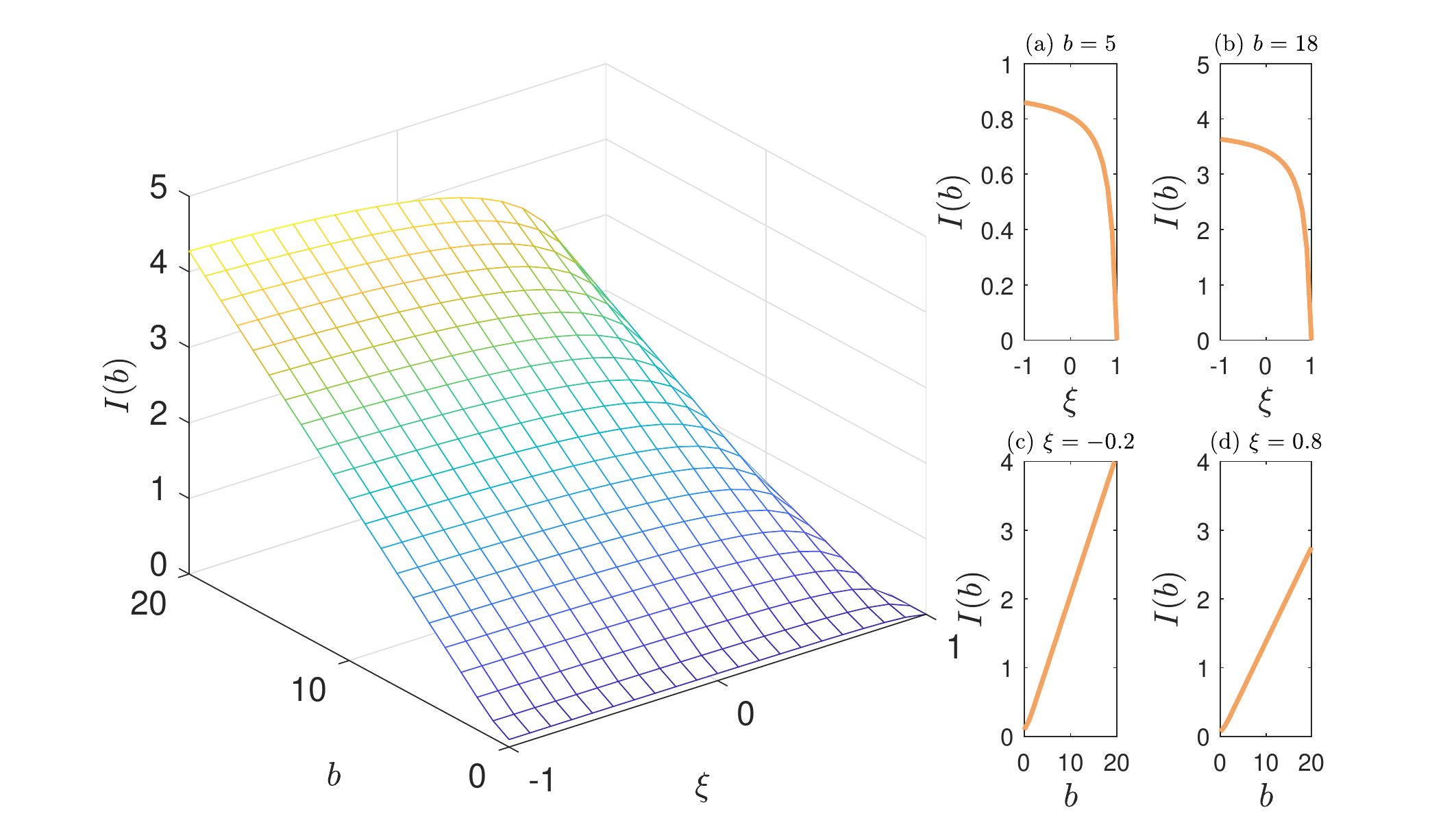}
\end{minipage}
}
\centering
\caption{The evolutions of $I(b)$ when $k=0.5$ and 2.}
\label{fig:I(b)2}
\end{figure}

Subsequently, we proceed to abstract the value of $I(b)$ under all cases into one figure to further investigate its properties. Fig. \ref{fig:I(b)2} shows the trends of $I(b)$ as $\xi\in(-1,1)$ and $b\in[0,20]$ when $\tilde{\mu}-\tilde{\lambda}=k$ is fixed as $0.5$ and 2, respectively. We can draw a conclusion that $I(b)$ evolves both according to $b$ as well as the autoregressive coefficient $\xi$. In detail, the rise of $\xi$ triggers the decrease of $I(b)$, almost in a negative logarithmical way. That is to say, when $\xi$ goes up, the decay speed of $P(\tilde{Q}>\tilde{\Omega})$ drops. The underlying reason may be that when the difference between $\tilde{\mu}$ and $\tilde{\lambda}$ is fixed, the block creation rate stresses more than network dissemination rate in affecting $I(b)$ considering the increase of $\xi$. Additionally, it can be found that $I(b)$ rises when $b$ goes up nearly in a linear way. This result is similar to that of $I(q)$ in the above section and \cite{corking}, which indicates that the natural forking robustness related parameter $b$ shows a direct positive influence on the rate function of $P(\tilde{Q}>\tilde{\Omega})$ proportionally. The reason behind this is that a higher $b$ represents a higher tolerance of blockchain in terms of natural forking, resulting in a higher decay speed of  $P(\tilde{Q}>\tilde{\Omega})$.

\begin{figure}[tpb]
\centering
\subfigure[ $\xi=-0.2$]{
\begin{minipage}[t]{0.45\linewidth}
\centering
\includegraphics[width=1.1\textwidth]{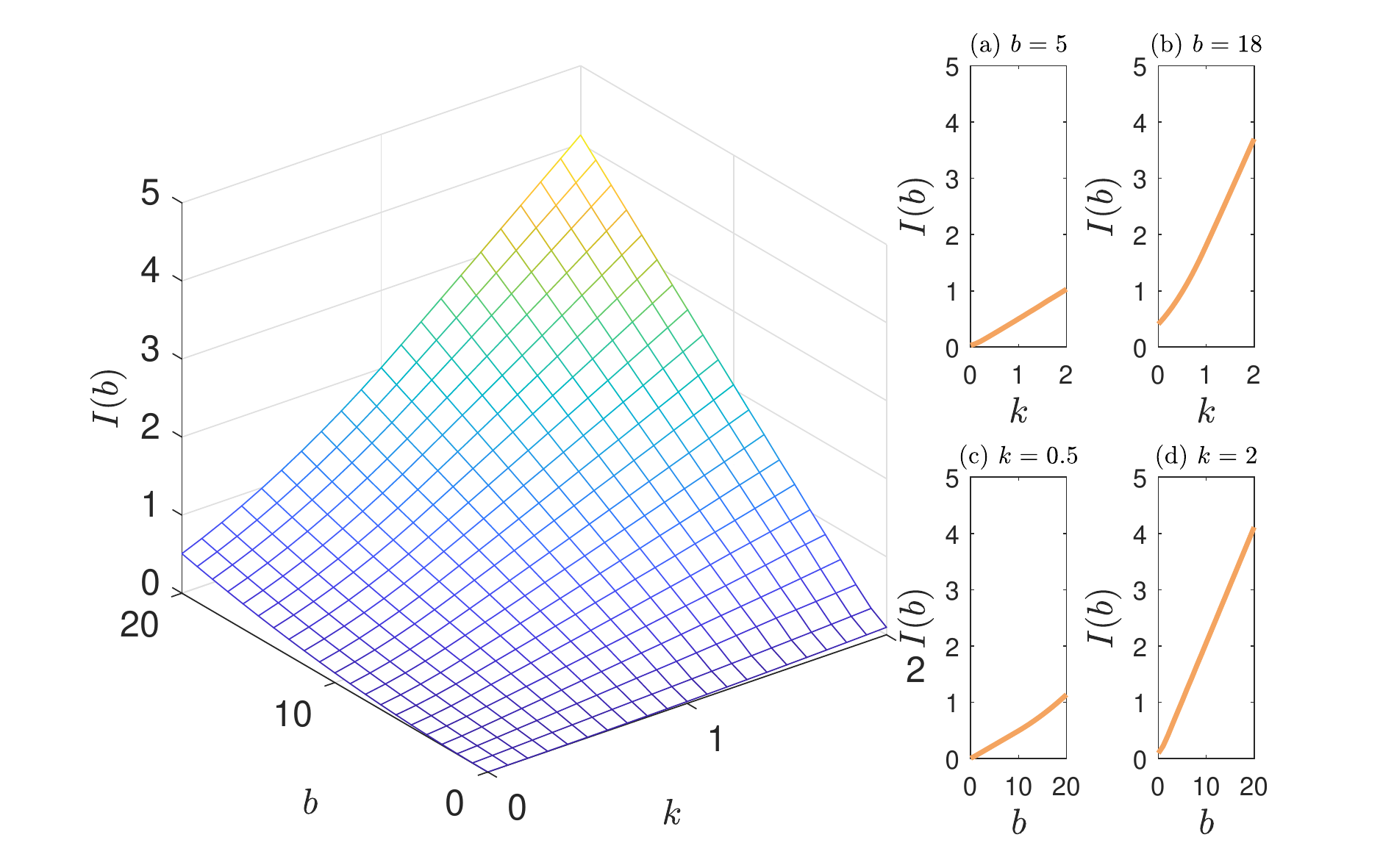}
\end{minipage}
}
\subfigure[$\xi=0.8$]{
\begin{minipage}[t]{0.45\linewidth}
\centering
\includegraphics[width=1.1\textwidth]{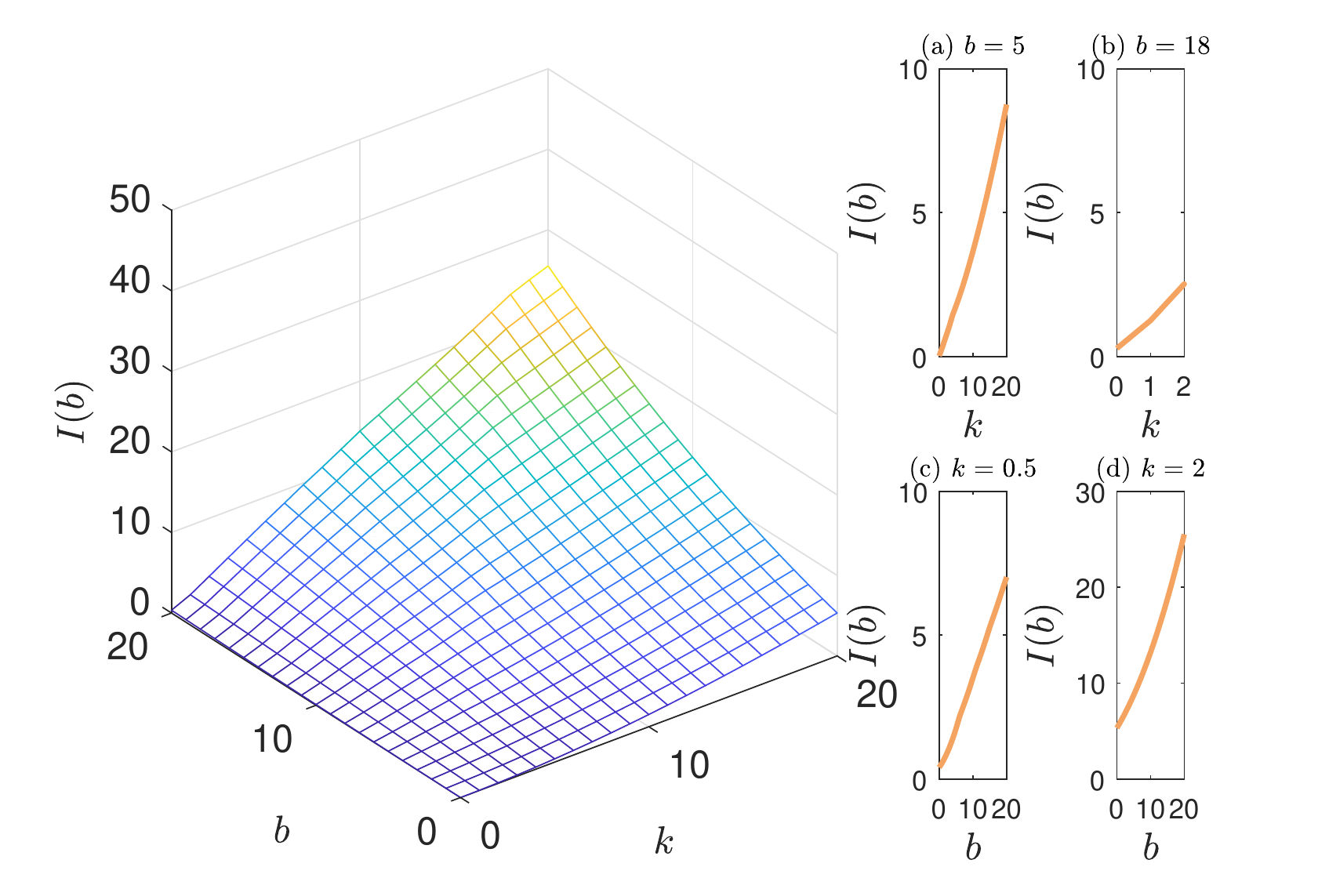}
\end{minipage}
}
\centering
\caption{The evolutions of $I(b)$ when $\xi=-0.2$ and 0.8.}
\label{fig:I(b)3}
\end{figure}

\begin{figure}[t]
\centering
\subfigure[ $b=5$]{
\begin{minipage}[t]{0.45\linewidth}
\centering
\includegraphics[width=1.1\textwidth]{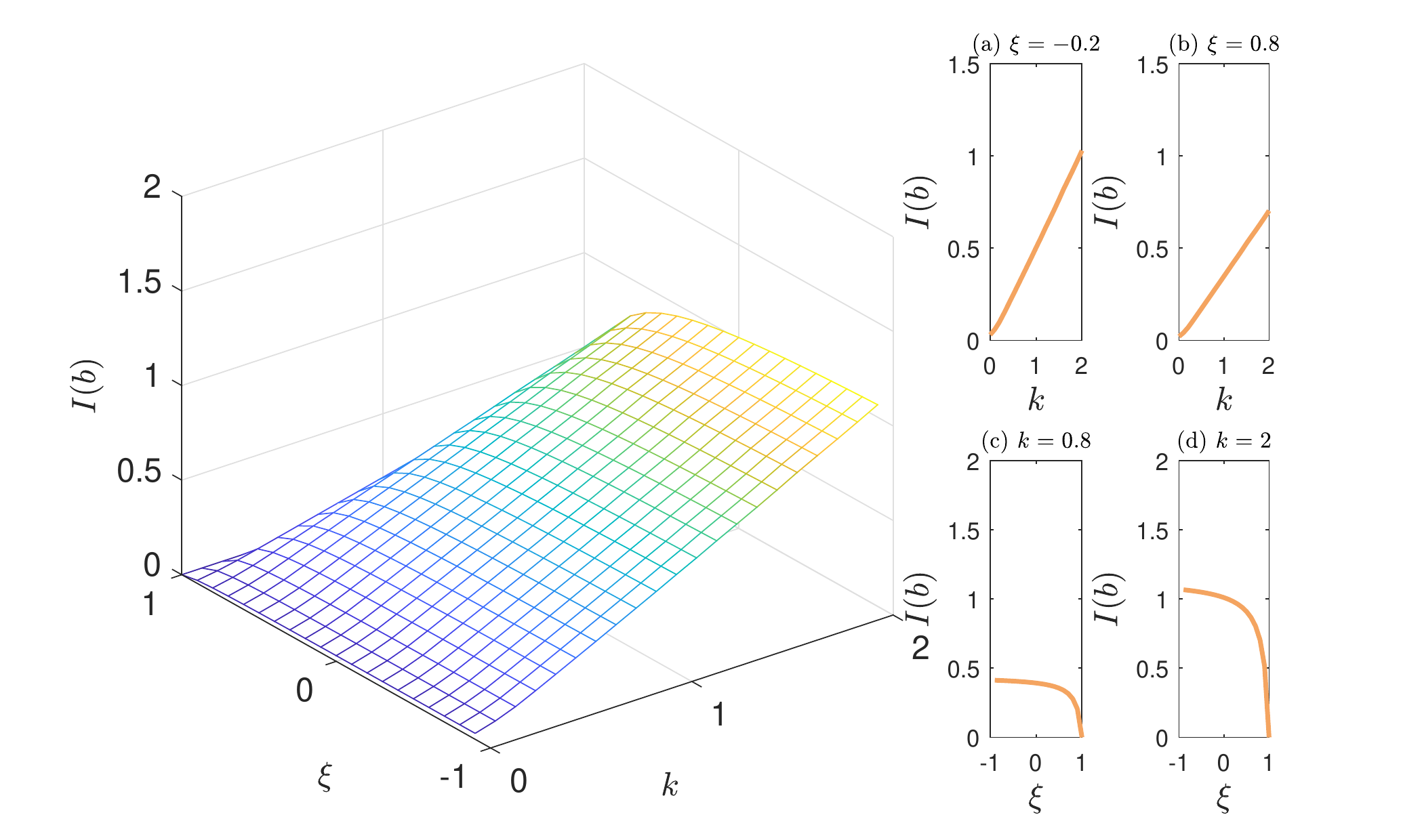}
\end{minipage}
}
\subfigure[ $b=10$]{
\begin{minipage}[t]{0.45\linewidth}
\centering
\includegraphics[width=1.1\textwidth]{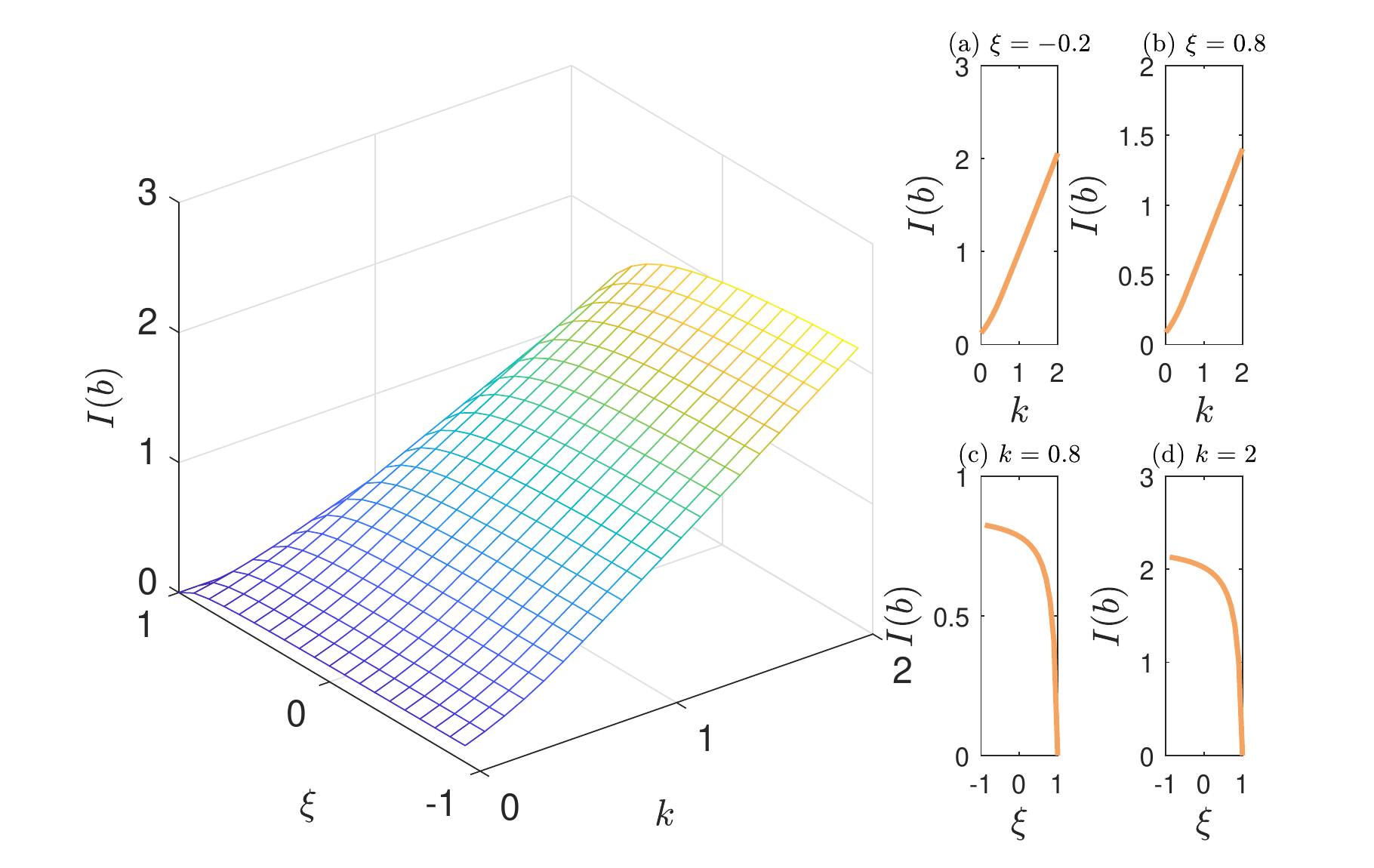}
\end{minipage}
}
\centering
\caption{The evolutions of $I(b)$ when $b=5$ and 10.}
\label{fig:I(b)4}
\end{figure}

Figs. \ref{fig:I(b)3} and \ref{fig:I(b)4} present the tendencies of $I(b)$ with respect to the defined $\xi=-0.2, 0.8$ and $b=5, 10$. According to these pictures, one more conclusion can be drawn that the increase of $k$, i.e., $\tilde{\mu}-\tilde{\lambda}$, can introduce the enhancement of $I(b)$ in an approximately linear way, resulting in quickly dropping of $P(\tilde{Q}>\tilde{\Omega})$. This can be interpreted in the following way: recall that $\tilde{\mu}-\tilde{\lambda}$ expresses the excessive amount of the network capability of transmitting blocks over that of block generation. If such an amount gets higher (i.e., a higher $k$), it is intuitive that the blockchain system is less exposed to natural forking.

Therefore, based on the parameter simulations carried out above, we can derive $I(b)$ informally as $I(b)=\tau(b)\cdot\upsilon(\xi)\cdot\omega(k)$, in which $\tau(b)$ and $\omega(k)$ respectively represent nearly positive linear relationships between $b,k$ and $I(b)$ while $\upsilon(\xi)$ denotes that $\xi$ shares a negative logarithmical connection with $I(b)$ approximately.
\section{Many-Source non-I.I.D. Scheme}\label{sec:duoyuan}

In this section, we relax the first assumption described in Section \ref{sec:danyuan non-iid} via treating each node that generates blocks as a single input. That is, a many-source non-i.i.d. scheme is considered, which enables a fine-grained analysis in natural forking complementary to the above two sections.

Define $\gamma_i^r$ and $\alpha_i^r=(N-1)\cdot\gamma_i^r$ as the number of mined blocks from source $r$  during the $i^{th}$ period and that of blocks required to be disseminated of $r$ in this interval. After that, let $\mathcal{A}_i^{N}$ be the number of blocks that need to be propagated from all nodes in the $i^{th}$ interval. That is, $\mathcal{A}_i^{N}=\alpha_i^1+...+\alpha_i^r+...+\alpha_i^N$ and we denote $\mathcal{S}_t^{N}=\mathcal{A}_0^{N}+...+\mathcal{A}_i^{N}+...+\mathcal{A}_t^{N}$ as the cumulative blocks that require to be disseminated for all the sources at the end of the $t^{th}$ period with expectation $\bar{\lambda}$ and $\mathcal{S}_0^{N}=0$. Accordingly, the difference in numbers of block created and broadcasted at the end of time $T$ can be given by
\begin{equation}
\begin{aligned}
\mathcal{Q}_t^{N}=\mathcal{S}_t^{N}-\mathcal{B}_t,
\end{aligned}
\end{equation}
where $\mathcal{B}_t=\Sigma_{j=1}^t\beta_j$ shows the network summing capability of transferring blocks until $T$ with expectation $\bar{\mu}>0$ and $\beta_j$ denotes the corresponding transferring capability in the $j^{th}$ period. Similarly, we focus on the maximum value of $\mathcal{Q}_t^{N}$, i.e., $\mathcal{Q}=\sup_{t\ge0}\mathcal{Q}_t^{N}$, to investigate the worst case of being threatened by natural forking when $t\to\infty$. Given the inconsistency-resistance degree $\bar{\Omega}$, the probability of $\mathcal{Q}>\bar{\Omega}$ reflects the threatened level of the blockchain system occurring natural forking, which presents a specific clue in striking back against this bifurcation.

It is worth noting that we consider the mining process under this scenario ($\mathcal{A}_i^{N}$) as a stationary Gaussian process instead of the Poisson process as mentioned in Section \ref{sec:danyuan iid} to depict the continuous setting of blockchain more practically and we allow $\{\alpha_i^r, i\in[1,t]\}$ to be a time-correlated process. Besides, we can infer that under the many-source non-i.i.d. scheme, the nodes may share equal computational power. That is to say, if we define $\mathcal{S}_t^r=\mathcal{A}_0^{r}+...+\mathcal{A}_i^{r}+...+\mathcal{A}_t^{r}$ to describe the cumulative number of blocks that need to be ``served" of source $r$ until time $T$, then we have $\mathbf{E}[S_t^1]=\mathbf{E}[S_t^2]=...=\mathbf{E}[S_t^N]$. This is because our considered many-source non-i.i.d. scheme indicates the case where the nodes' computing powers are involved and entangled with each other, which suggests the emergence of {\it collusion}. And only when the computing powers among nodes are the same will incur the phenomenon of collusion since there is no motivation for a node that possesses a higher computing power to share its mining advantages with someone with lower computing power.

In the following, before presenting the formal deviations of $P(\mathcal{Q}>\bar{\Omega})$, we will first propose several basic assumptions for a many-source non-i.i.d. queue to further inspect its intrinsic properties according to \cite{big} and \cite{wangxia}.

\begin{assumption}\label{jiashe1}
Define $\bar{\Lambda}_{N,t}^\mathcal{S}(\theta)=\log \mathbf{E}[e^{\theta \mathcal{S}_t^N}]$, and $\forall N,t$, $\bar{\Lambda}_{N,t}^\mathcal{S}(\theta)$ is finite for $\theta$ in a neighborhood of the origin.
\end{assumption}

\begin{assumption}\label{jiashe2}
Suppose $\forall \theta\in \mathbb{R}$, limit $\bar{\Lambda}_t^\mathcal{S}(\theta)=\lim_{N\to\infty}\frac{1}{N}\bar{\Lambda}_{N,t}^\mathcal{S}(\theta)$ exists and $\bar{\Lambda}_t^\mathcal{S}(\theta)\in\mathbb{R}$. Besides, $\bar{\Lambda}_t^\mathcal{S}(\theta)$ is finite in a neighborhood of $\theta=0$ and essentially smooth and lower-semicontinuous.
\end{assumption}

\begin{assumption}\label{jiashe3}
Suppose there is a sequence $(v_t,t\in\mathbb{N})$ taking values in $\mathbb{R^+}$ and satisfying $\frac{v_t}{\log t}\to\infty$. And $\forall \theta$, the following equation holds:
\begin{equation}\label{jiashegongshi}
\lim_{N\to\infty}\sup_t|\frac{1}{N}\log\mathbf{E}[e^{\theta v_t \mathcal{S}_t^N/t}]-\bar{\Lambda}_t^\mathcal{S}(\theta v_t/t)|=0.
\end{equation}
\end{assumption}

\begin{assumption}\label{jiashe4}
Suppose the cumulant generating function of $\mathcal{S}_t^{N}$, i.e., $\bar{\Lambda}_\mathcal{S}(\theta)=\lim_{t\to\infty}\frac{\bar{\Lambda}_t^\mathcal{S}(\theta v_t/t)}{v_t}$, exists and  $\bar{\Lambda}_\mathcal{S}(\theta)\in\mathbb{R}$. Besides, it is finite and differentiable in a neighborhood of the origin.
\end{assumption}

Under the assumptions of  \ref{jiashe1}-\ref{jiashe4}, conclusions similar to \eqref{p2} and \eqref{I(b)} may still hold for the many-source non-i.i.d. scheme in light of the generalized Cramer's theorem \cite{big}. More specifically, if $\bar{\Omega}=N\cdot u$ with $N\to\infty$, then $P(\mathcal{Q}>\bar{\Omega})$ can be calculated analogously by
\begin{equation}\label{disanzhangI}
\begin{aligned}
\lim_{N\to\infty}\frac{1}{N}\log P(\mathcal{Q}>\tilde{\Omega})=-I(u),
\end{aligned}
\end{equation}
where $I(u)$ is described as
\begin{equation}\label{I(u)}
\begin{aligned}
I(u)=\mathop{\inf}_{t>0}t\cdot\bar{\Lambda}^*(\frac{u}{t}).
\end{aligned}
\end{equation}

In \eqref{I(u)}, $\bar{\Lambda}^*(\bar{x})$ denotes the convex conjugate of $\mathcal{Q}_t^N$ taking substitution of $\frac{u}{t}$ as $\bar{x}$. Similar to the description in the previous sections, $\bar{\Lambda}^*(\bar{x})$ can be derived through $\bar{\Lambda}^*(\bar{x})=
\sup_{\theta\in\mathbb{R}}\{\theta \bar{x}-\bar{\Lambda}_\mathcal{S}(\theta)-\bar{\Lambda}_\mathcal{B}(-\theta)\}=\mathop{\inf}_{\bar{y}\in\mathbb{R}}\{\bar{\Lambda}^*_\mathcal{S}(\bar{y})+\bar{\Lambda}^*_\mathcal{B}(\bar{y}-\bar{x})\}, \forall \bar{x}\in\mathbb{R}$. Since the presentation  of $\mathcal{B}_t$ is comparable to that of $B_t$ and $\tilde{B_t}$, we then omit to present the derivation of $\bar{\Lambda}^*_\mathcal{B}(\cdot)$. Nonetheless, $\mathcal{S}_t^N$ is defined differently as a stationary Gaussian process. Hence, to deduce $\bar{\Lambda}_\mathcal{S}(\theta)$ and $\bar{\Lambda}^*(\cdot)$ formally, we first introduce the following definition and assumption.

\begin{definition}(Stationary Gaussian process)\label{gaussian}
Let the average input of the many-source non-i.i.d. queue follows a stationary Gaussian distribution, which can be characterized by the following structure:
$$
(\frac{1}{N}\mathcal{A}_0^N,...,\frac{1}{N}\mathcal{A}_t^N)'\sim Normal(\bar\lambda \epsilon,\Sigma_t),
$$
where $\bar\lambda \epsilon$ and $\Sigma_t$ represent the mean and covariance with $\epsilon=(1,...,1)'_{1\times t}$ and $(\Sigma_t)_{ij}=\mathbf{Cov}(\frac{1}{N}\mathcal{A}_i^N,\frac{1}{N}\mathcal{A}_j^N)$.
\end{definition}

\begin{assumption}\label{jiashe5}
Let $\Upsilon (N)=\mathbf{Var}\frac{1}{N}\mathcal{S}_t^N$ and suppose $\lim_{N\to\infty}N\cdot \Upsilon (N)$ exists, denoted as $\varsigma(t)$. Besides, presume there is a sequence $(v_t, t\in\mathbb{N})$ taking values in $\mathbb{R}^+$ and satisfying $\frac{v_t}{\log t}\to\infty$ with the existence of $\lim_{t\to\infty}\frac{v_t\varsigma(t)}{t^2}$ (termed as $\mathbcal{c}$), and
\begin{equation}\label{proof}
\lim_{N\to\infty}\sup_t|\frac{v_t^2}{t^2}N\Upsilon (N)-\frac{v_t^2}{t^2}\varsigma(t)|=0.
\end{equation}
\end{assumption}

Based on the above definition and assumption, we present the following theorem to demonstrate $\bar{\Lambda}_{\mathcal{S}}(\theta)$ formally, paving the way for deriving $I(u)$ subsequently.

\begin{theorem}\label{manysources}
If the inputs of the many-source non-i.i.d. scheme are stated as Definition \ref{gaussian}, when satisfying assumption \ref{jiashe5}, the cumulant generating function of $\mathcal{S}_t^N$, i.e.,
$\bar{\Lambda}_{\mathcal{S}}(\theta)$, can be derived by
\begin{equation}\label{cumulant}
\bar{\Lambda}_{\mathcal{S}}(\theta)=\theta\bar{\lambda}+\frac{1}{2}\theta^2.
\end{equation}
\end{theorem}

\begin{IEEEproof}
To start with, we corroborate that assumption \ref{jiashe5} is sufficient to meet assumptions \ref{jiashe1}-\ref{jiashe4} to deduce $\bar{\Lambda}_{\mathcal{S}}(\theta)$. Recall that
$\bar{\Lambda}_{N,t}^\mathcal{S}(\theta)$ is defined as $\bar{\Lambda}_{N,t}^\mathcal{S}(\theta)=\log \mathbf{E}[e^{\theta \mathcal{S}_t^N}]$, which can be further calculated as $\log \mathbf{E}[e^{\theta (\mathcal{A}_0^N+\cdot+\mathcal{A}_t^N)}]
=\theta N \bar{\lambda} t+\frac{1}{2} \theta^2 N^2 \Upsilon(N).$ It is obvious that $\forall N,t$, $\bar{\Lambda}_{N,t}^\mathcal{S}(\theta)$ is finite for $\theta$ in a neighborhood of the origin, indicating that assumption \ref{jiashe1} could be met in our case. Subsequently, we take the limit as $N\to\infty$ of $\bar{\Lambda}_{N,t}^\mathcal{S}(\theta)$, which yields $\bar{\Lambda}_t^\mathcal{S}(\theta)=\lim_{N\to\infty}\frac{1}{N}\bar{\Lambda}_{N,t}^\mathcal{S}(\theta)
=\theta\bar{\lambda}t+\frac{1}{2}\theta^2\varsigma(t)$ because of $\lim_{N\to\infty}N\cdot \Upsilon(N)=\varsigma(t)$. Hence, the limit $\bar{\Lambda}_{N,t}^\mathcal{S}(\theta)$ exists and is finite on a neighborhood of $\theta=0$. Moreover, $\bar{\Lambda}_{t}^\mathcal{S}(\theta)$ is essentially smooth and lower-semicontinuous since (1) for a sequence $(\theta_n, n\in \mathbb{R}^+)$, $\lim_{n\to +\infty}|\nabla\bar{\Lambda}_{t}^\mathcal{S}(\theta_n)|=+\infty$ and (2) all the level sets of $\bar{\Lambda}_{t}^\mathcal{S}(\theta)$ are closed where the level set is determined by $\{\theta:\bar{\Lambda}_{t}^\mathcal{S}(\theta)\le\mathcal{a}\}$ for $\mathcal{a}\in \mathbb{R}$ as demonstrated in \cite{big}. This testifies that assumption \ref{jiashe2} could be satisfied.

As for assumption \ref{jiashe3}, we proceed to transform \eqref{jiashegongshi} in a more intuitive way. Let $\mathcal{f}=\frac{1}{N}\log\mathbf{E}[e^{\theta v_t \mathcal{S}_t^N/t}]-\bar{\Lambda}_t^\mathcal{S}(\theta v_t/t)$. On account of $\log \mathbf{E}[e^{\theta \mathcal{S}_t^N}]=\theta N \bar{\lambda} t+\frac{1}{2} \theta^2 N^2 \Upsilon(N)$ and $\bar{\Lambda}_t^\mathcal{S}(\theta)=\theta\bar{\lambda}t+\frac{1}{2}\theta^2\varsigma(t)$, we can find that
\begin{equation}\label{jiashegongshi2}
\begin{aligned}
\mathcal{f}&=\frac{1}{N}[\theta \frac{v_t}{t} N\bar{\lambda}t+\frac{1}{2}\frac{\theta^2v_t^2}{t^2}N^2\Upsilon(N)]-[\theta \frac{v_t}{t}\bar{\lambda}t+\frac{1}{2}\frac{\theta^2v_t^2}{t^2}\varsigma(t)]\\
&=\frac{1}{2}\frac{\theta^2v_t^2}{t^2}N\Upsilon(N)-\frac{1}{2}\frac{\theta^2v_t^2}{t^2}\varsigma(t).
\end{aligned}
\end{equation}
Therefore, in order to meet \eqref{jiashegongshi}, it is required to satisfy $\lim_{N\to\infty}\sup_t|\frac{v_t^2}{t^2}N\Upsilon(N)-\frac{v_t^2}{t^2}\varsigma(t)|=0$ according to \eqref{jiashegongshi2} without taking the coefficient $\frac{1}{2}$ and irrelevant parameter $\theta$ into consideration. As a result, we state that assumption \ref{jiashe3} can be met. In addition, if there is a sequence $(v_t,t\in\mathbb{N})$ which takes values in $\mathbb{R^+}$ with $\frac{v_t}{\log t}\to\infty$, then the cumulant generating function $\bar{\Lambda}_S(\theta)$ can be deduced as $\bar{\Lambda}_S(\theta)=\lim_{t\to\infty}\frac{\theta v_t \bar{\lambda}+\frac{\theta^2 v_t^2 \varsigma(t)}{2t^2}}{v_t}=\theta\bar{\lambda}+\lim_{t\to\infty}\frac{\theta^2v_t\varsigma(t)}{2t^2}$. Evidently, we can obtain that $\bar{\Lambda}_S(\theta)=\theta\bar{\lambda}+\frac{1}{2}\theta^2\mathbcal{c}$ due to $\lim_{t\to\infty}\frac{v_t\varsigma(t)}{t^2}=\mathcal{c}$. Hence, the existence of $\bar{\Lambda}_S(\theta)$ is proved and it is finite and differentiable in a neighborhood of the origin, convincing the validation of assumption \ref{jiashe4}.

Now we are in a position to prove equation \eqref{cumulant}. In the above paragraph, it is derived that $\bar{\Lambda}_S(\theta)=\theta\bar{\lambda}+\frac{1}{2}\theta^2\mathbcal{c}$, where $\mathcal{c}$ is defined as $\lim_{t\to\infty}\frac{v_t\varsigma(t)}{t^2}$. In order to further demonstrate $\bar{\Lambda}_S(\theta)$, here we clarify the meaning of $v_t$ more concretely. Note that $v_t$ is proposed in assumption \ref{jiashe3} to ensure the existence of $\bar{\Lambda}_\mathcal{S}(\theta)$ when taking limit $t\to\infty$ (assumption \ref{jiashe4}). This is because the nonexistence of $\bar{\Lambda}_\mathcal{S}(\theta)$ always occurs when distribution like the notably fractional Brownian motion is involved as presented in \cite{big}. For such distributions, Ganush {\it et al.} \cite{big} introduced the so-called {\it scaling function} $v_t$ to describe the correlations decaying with time. By doing so, when $t\to\infty$, the limit of $\frac{\bar{\Lambda}_t^\mathcal{S}(\theta v_t/t)}{v_t}$ will always exist. Particularly, $v_t$ can be defined in different forms serving for various problems.\footnote{For example, for sources distributed as a fractional Brownian motion as mentioned above, $v_t$ is often established as a power of $t$, regarding as the {\it power-law scalings}.} Aware of this, the natural choice of the scaling function $v_t$ in our case is $v_t=\frac{t^2}{\varsigma(t)}$ for simplicity. As a result, $\mathcal{c}=1$ and \eqref{cumulant} is finally proved.
\end{IEEEproof}

\begin{figure}[t]
\centering
\subfigure[$h=2$]{
\begin{minipage}[t]{0.45\linewidth}
\centering
\includegraphics[width=1.1\textwidth]{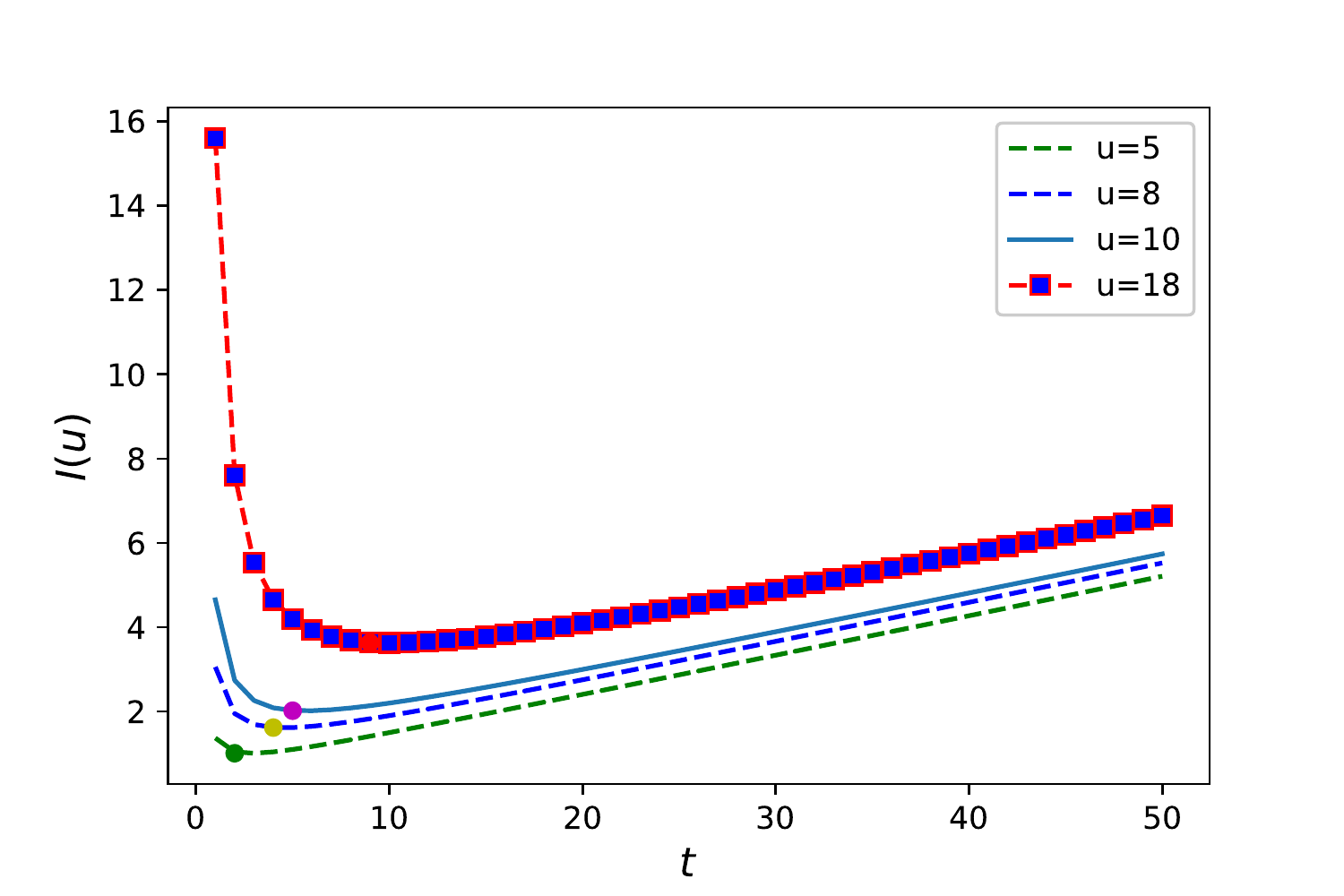}
\end{minipage}
}
\subfigure[ $h=5$]{
\begin{minipage}[t]{0.45\linewidth}
\centering
\includegraphics[width=1.1\textwidth]{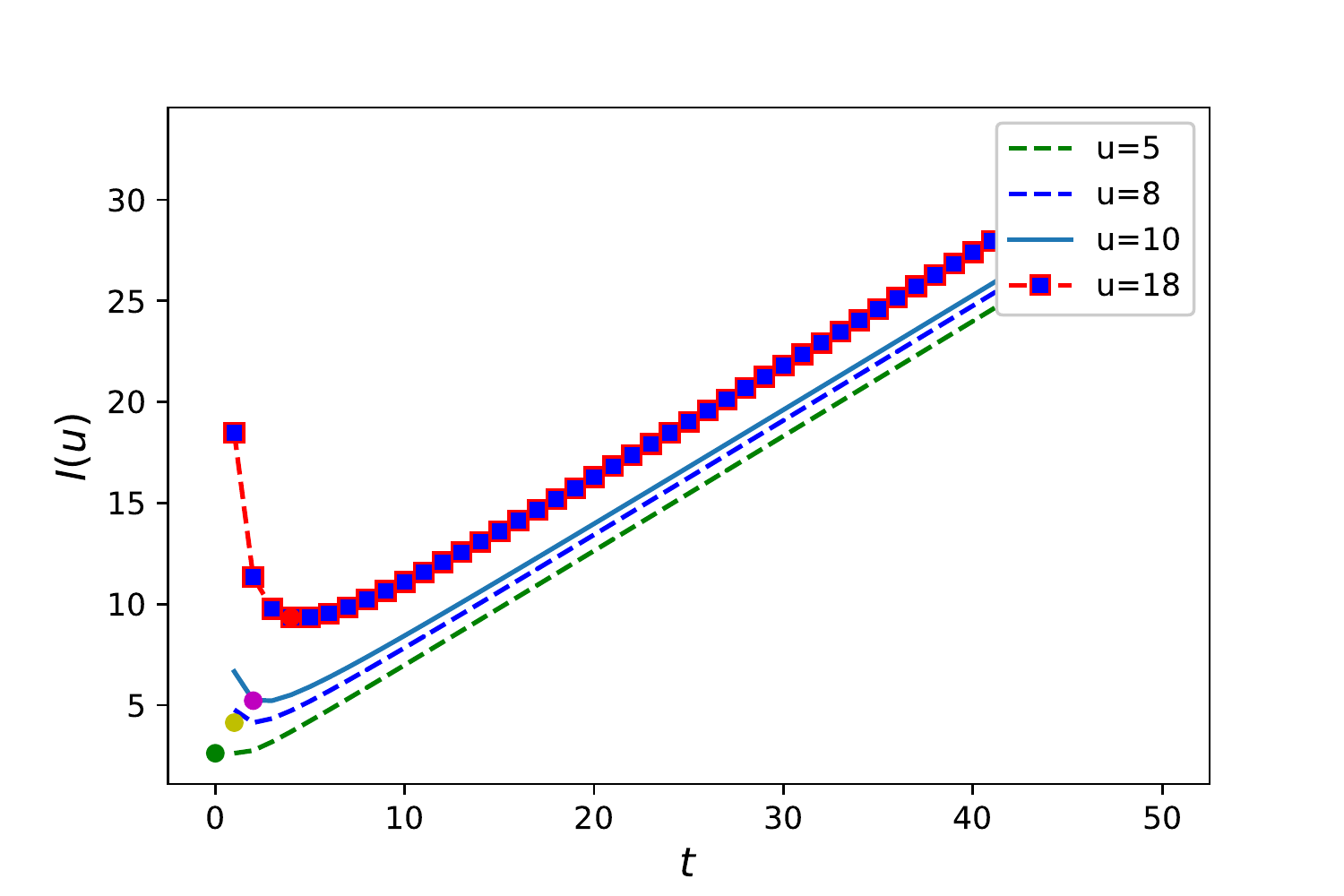}
\end{minipage}
}
\centering
\caption{The evolutions of $I(u)$ when $h=2$ and 5.}
\label{fig:I(u)1}
\end{figure}

\begin{figure}[t]
\centering
\subfigure[ $u=5$]{
\begin{minipage}[t]{0.45\linewidth}
\centering
\includegraphics[width=1.1\textwidth]{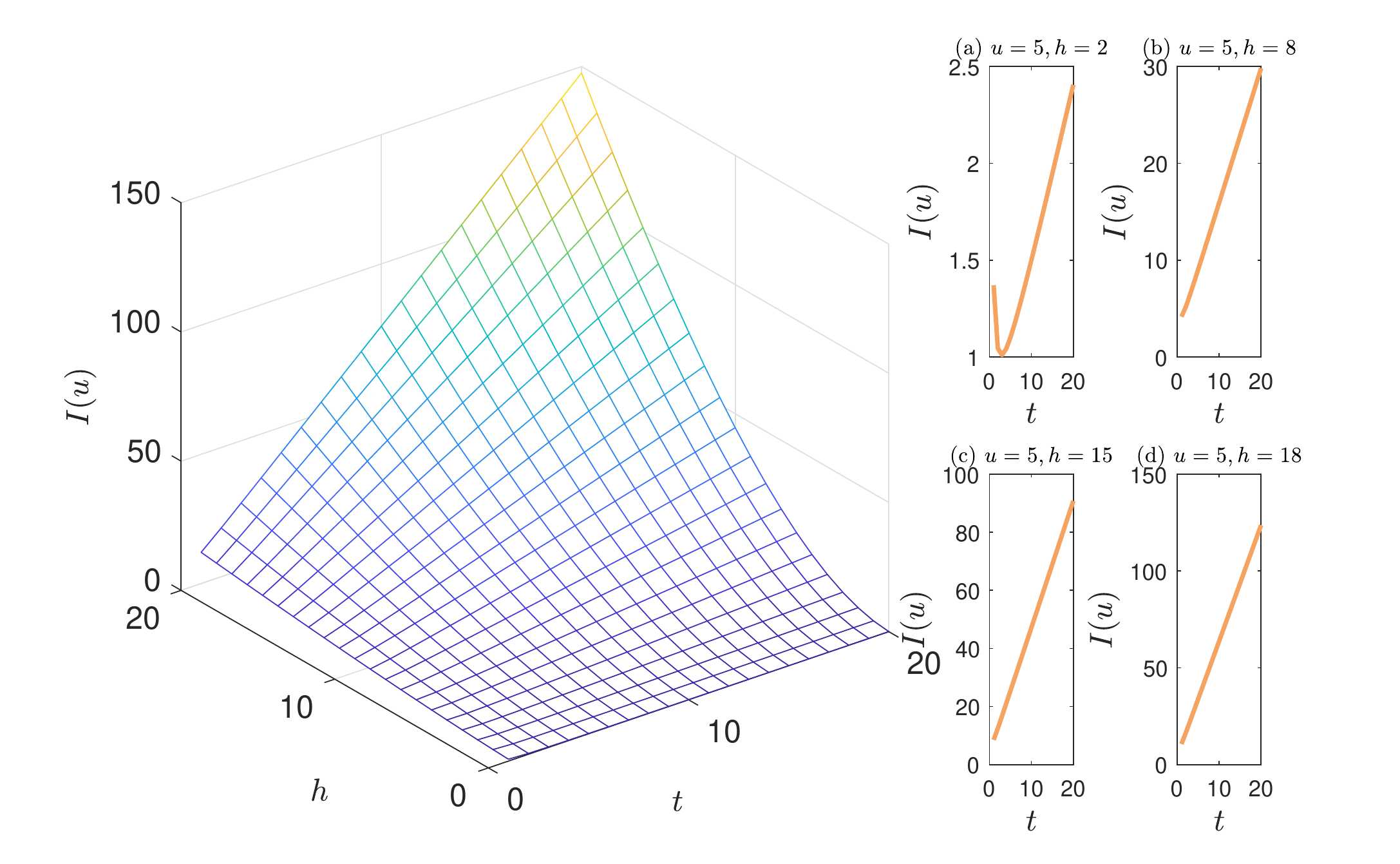}
\end{minipage}
}
\subfigure[ $u=8$]{
\begin{minipage}[t]{0.45\linewidth}
\centering
\includegraphics[width=1.1\textwidth]{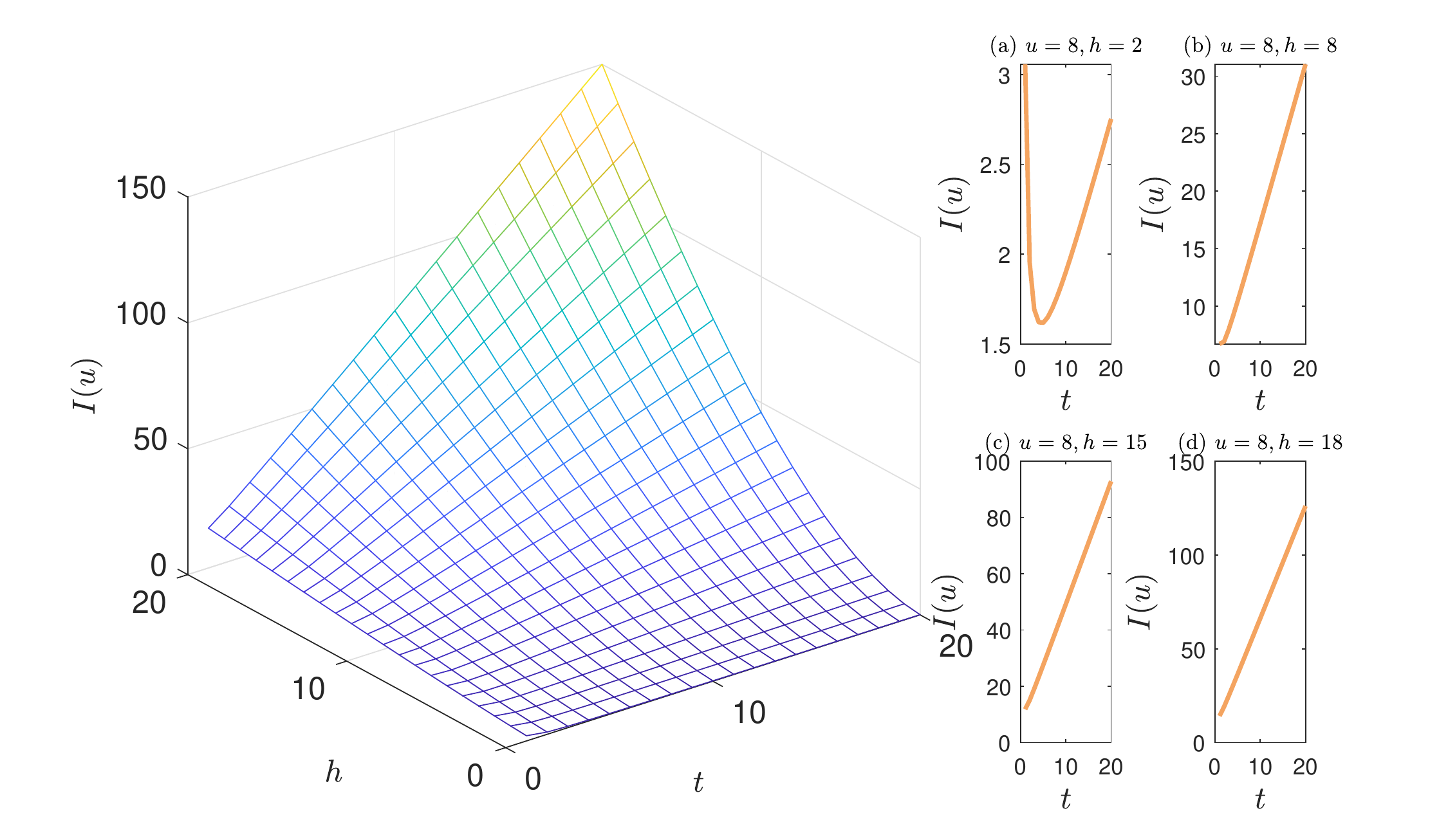}
\end{minipage}
}
\centering
\caption{The evolutions of $I(u)$ when $u=5$ and 8.}
\label{fig:I(u)2}
\end{figure}

\begin{figure}[t]
\centering
\includegraphics[width=3in]{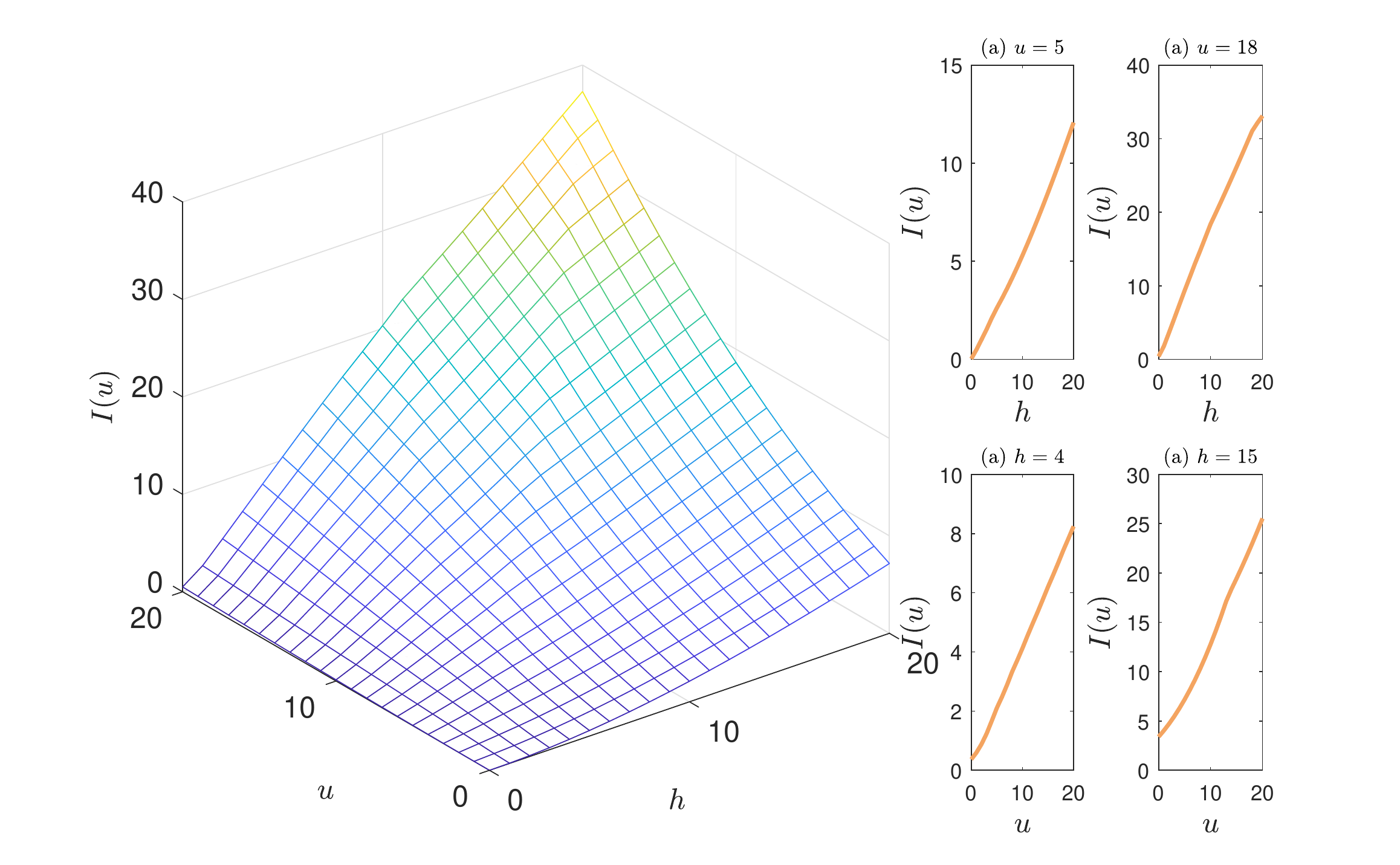}
\caption{The evolution of $I(u)$ in the many-source non-i.i.d. scheme.}
\label{fig:I(u)3}
\end{figure}

After obtaining $\bar{\Lambda}_{\mathcal{S}}(\theta)$ in Theorem \ref{manysources}, we present its corresponding convex conjugate $\bar{\Lambda}_{\mathcal{S}}^*(\bar{x})$ by calculating the supremum of  $\theta\bar{x}-\bar{\Lambda}_{\mathcal{S}}(\theta)$ when $\theta\in\mathbb{R}$. In detail, let $\mathcal{g}(\theta)=\theta\bar{x}-\bar{\Lambda}_{\mathcal{S}}(\theta)$, then the first- and second-order partial derivatives are $\frac{\partial \mathcal{g}(\theta)}{\partial \theta}=\bar{x}-\bar{\lambda}-\theta$ and $\frac{\partial^2 \mathcal{g}(\theta)}{\partial \theta^2}=-1$. Hence, we can get $\bar{\Lambda}_{\mathcal{S}}^*(\bar{x})=\frac{1}{2}(\bar{x}-\bar{\lambda})^2$ when $\theta=\bar{x}-\bar{\lambda}$. Considering that $\bar{\Lambda}_{\mathcal{B}}^*(\bar{x})$ is identical with $\Lambda^*_{B}(x)$ because of $\bar{\Lambda}_{\mathcal{B}}(\theta)=\Lambda_{B}(\theta)$, we have
$\bar{\Lambda}^*(\bar{x})=\mathop{\inf}_{\bar{y}\in\mathbb{R}}\{\bar{\Lambda}^*_\mathcal{S}(\bar{y})+\bar{\Lambda}^*_\mathcal{B}(\bar{y}-\bar{x})\}=\mathop{\inf}_{\bar{y}\in\mathbb{R}}\bar{\Gamma}(\bar{y}), \forall \bar{x}\in\mathbb{R}$
in which
\begin{equation}\label{chaoyue2}
\bar{\Gamma}(\bar{y})=\frac{(\bar{y}-\bar{\lambda})^2}{2}+(\bar{y}-\bar{x})\log\frac{\bar{y}-\bar{x}}{\bar{\mu}}+\bar{\mu}-(\bar{y}-\bar{x}).
\end{equation}

Therefore, in order to get the extreme point of $\bar{\Gamma}(\bar{y})$, namely, $\bar{\Gamma}(\bar{y})^*$, we need to attain the solution of the following transcendental equation:
\begin{equation}\label{chaoyue3}
(\bar{y}-\bar\lambda)+\log\frac{\bar{y}-\bar{x}}{\bar\mu}=0,
\end{equation}
when $\bar{y}>\bar{x}>0$. Then similar to \eqref{chaoyue}, we carry out Taylor Expansion of $\log\frac{\bar{y}-\bar{x}}{\bar\mu}$ to solve \eqref{chaoyue3}. Hence, $\log\frac{\bar{y}-\bar{x}}{\bar\mu}$ turns to $\frac{\bar{y}-\bar{x}-\bar{\mu}}{\bar{\mu}}-\frac{(\bar{y}-\bar{x}-\bar{\mu})^2}{2\bar{\mu}^2}$ and the analytical solutions of \eqref{chaoyue3} can be derived as $\bar{y_1}$ and $\bar{y_2}$, in which
\begin{equation}\label{solutions2}
\begin{cases}
\bar{y_1}=2\bar{\mu}+\bar{\mu}^2+\bar{x}-\bar{\mu}\varpi,\\
\bar{y_2}=2\bar{\mu}+\bar{\mu}^2+\bar{x}+\bar{\mu}\varpi,
\end{cases}
\end{equation}
with $\varpi=(1-2\bar{\lambda}+4\bar{\mu}+\bar{\mu}^2+2\bar{x})^{\frac{1}{2}}$. As a result, we need to replace $\bar{y}$ with $\bar{y_1}$ and $\bar{y_2}$ in \eqref{chaoyue2} to get $\bar{\Gamma}(\bar{y})^*$.  By doing so, the rate functions $I(u)$ can be calculated through obtaining the infimum of $t\cdot \bar{\Lambda^*}(\frac{u}{t})$ when $t>0$ based on \eqref{I(u)}.

Owing to the computational complexity in acquiring the analytical solution of $I(u)$ as demonstrated in Section \ref{sec:danyuan non-iid}, we resort to executing parameter simulations in the following to reveal some key insights of the decay rate of $P(\mathcal{Q}>\tilde{\Omega})$. The parameter settings and experimental requirements employed in this section are comparable to those of the previous section, which are briefly depicted as: 1) $\bar{y}=\bar{y_1}$ in \eqref{chaoyue2} instead of $\bar{y}=\bar{y_2}$ due to the nonexistence of infimum of $\bar{\Gamma}(\bar{y_2})$; 2) set $\bar{\lambda}>u$ so as to meet the requirement of $\bar{y}>\bar{x}>0$; 3) although we have testified numerous experiments when $t$ goes very large, i.e., $t\to2000$, and under different parameter settings (i.e., $h=\bar{\mu}-\bar{\lambda}\in[0,2000]$ and $u\in[0,500]$), only parts of the results are presented in our paper since they exhibit similar trends and thus are omitted considering the page limitation.

In the following, we conduct numerical simulations comparable to the previous section. Firstly, since $I(u)$ presents the infimum of $t\cdot \bar{\Lambda^*}(\frac{u}{t})$, it is required to corroborate that $I(u)$ possesses such infimum when $t$ varies, which are depicted in Fig. \ref{fig:I(u)1} and Fig. \ref{fig:I(u)2}. Detailedly, Fig. \ref{fig:I(u)1} exhibits the trends of $I(u)$ with respect to $h=2,5$ and $u=5,8,10,18$. As can be seen from subfigures (a) and (b), each line decreases first and increases subsequently, indicating the existence of infimum in each case. Additionally, Fig. \ref{fig:I(u)2} expresses the tendencies of $I(u)$ in a three-dimensional way when $t$ and $h$ vary in $[0,20]$ with $u=5,8$, respectively. The subfigures (a) (b) also prove that there is infimum of $I(u)$ under each case.

After testifying the existence of $I(u)$, we are going to investigate its properties with regard to different arguments, as shown in Fig. \ref{fig:I(u)3}. Fig. \ref{fig:I(u)3} reports the evolution of $I(u)$ as $u$ and $h$ belonging to $[0,20]$. Obviously, we can conclude that the rise of $u$ will incur the increase of $I(u)$ nearly in a linear way and so does $h$. This is intuitive to understand since 1) a higher $u$ represents the blockchain system is more robust in enduring natural forking, and as a result, the decay rate of $P(\mathcal{Q}>\tilde{\Omega})$ becomes greater in this case; 2) a higher $h$ indicates that the network transmission capacity is much greater than the block completion rate, which can dredge the backlog of blocks quicker, making less room for natural forking. Based on the parameter simulations carried out above, we can derive $I(u)$ implicitly as $I(u) =\eta(u)\cdot\zeta(h)$, where $\eta(u)$ and $\zeta(h)$ respectively express the positive linear relationships between $u,h$ and $I(b)$ approximately.

\section{Conclusion}\label{sec:conclusion}
In this paper, we analyzed natural forking in blockchain from a {\it micro} point of view by investigating the {\it instantaneous difference} between block generation and dissemination. To that aim, we first construct a queuing model and propose the concept of {\it inconsistency-resistance degree}, based on which, we leverage the large deviation theory to study the natural forking probability and its decay rate in three cases comprehensively and complementarily. In addition, we bring about the concepts of {\it effective inconsistency-resistance degree} and {\it effective network transmission rate}, whose solutions facilitate to design defensive mechanisms for technically and operatively resisting natural forking of blockchain. According to our  solid theoretical derivation and extensive numerical simulations, we find: 1) the probability of the mismatch between  block generation and dissemination exceeding a given threshold dwindles exponentially with the increase of natural forking robustness related parameter or the difference between the block dissemination rate and the block creation rate; 2) the natural forking robustness related parameter may emphasize a more dominant effect on accelerating the abortion of natural forking in some cases; 3) when the self-correlated block generation rate is depicted as the stationary autoregressive process with a scaling parameter, it is found that setting a lower scaling parameter may speed up the failure of natural forking. These valuable findings have laid theoretical foundations to boost meticulous understanding of natural forking, providing specific clues to devise optimal countermeasures for thwarting natural forking.

\ifCLASSOPTIONcompsoc
  \section*{Acknowledgments}
\else
  \section*{Acknowledgment}
\fi
This work has been supported by National Key R$\&$D Program of China (No. 2019YFB2102600), National Natural Science Foundation of China (No. 61772080 and 62072044), the Blockchain Core Technology Strategic Research Program of Ministry of Education of China (No. 2020KJ010301), BNU Interdisciplinary Research Foundation for the First-Year Doctoral Candidates (No. BNUXKJC2022), the International Joint Research Project of Faculty of Education, Beijing Normal University, and Engineering Research Center of Intelligent Technology and Educational Application, Ministry of Education.

\ifCLASSOPTIONcaptionsoff
  \newpage
\fi

\bibliographystyle{IEEEtran}
\bibliography{arxiv}

\begin{IEEEbiography}[{\includegraphics[width=1in,height=1.25in,clip,keepaspectratio]{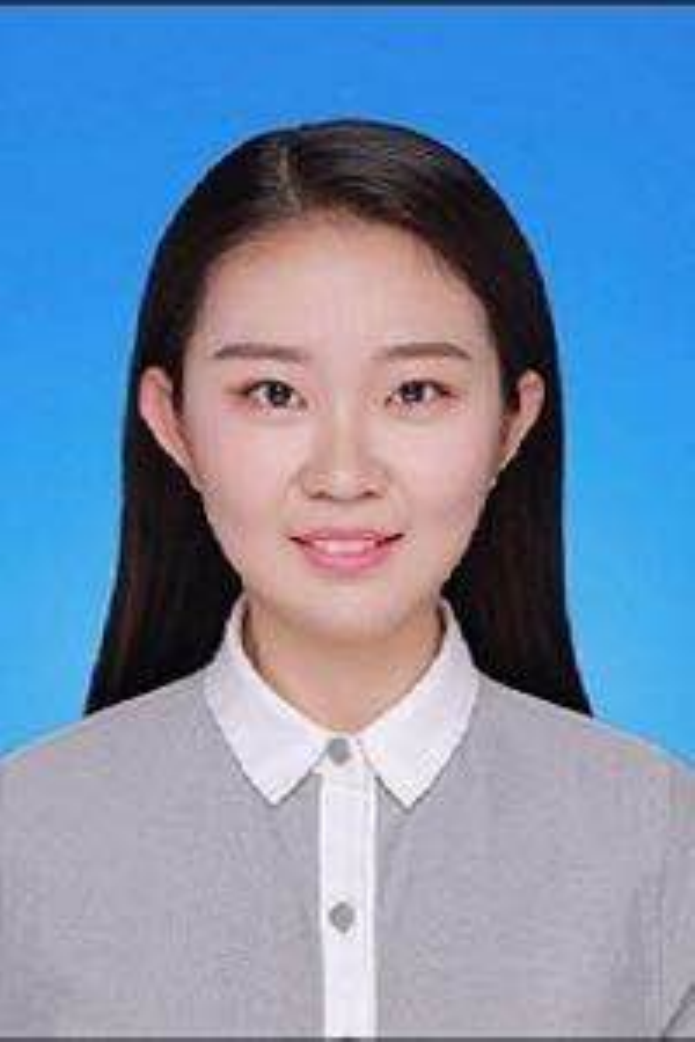}}]
{Hongwei Shi}received her B.S. degree in Computer Science from Beijing Normal University in 2018. Now she is pursuing her Ph.D. degree in Computer Science from Beijing Normal University. Her research interests include blockchain, game theory and combinatorial optimization.
\end{IEEEbiography}

\begin{IEEEbiography}[{\includegraphics[width=1in,height=1.25in,clip,keepaspectratio]{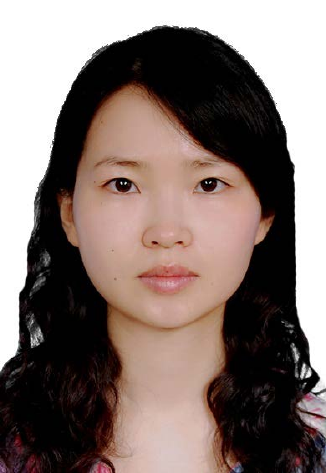}}]{Shengling Wang} is a full professor in the School of Artificial Intelligence, Beijing Normal University. She received her Ph.D. in 2008 from Xi'an Jiaotong University. After that, she did her postdoctoral research in the Department of Computer Science and Technology, Tsinghua University. Then she worked as an assistant and associate professor from 2010 to 2013 in the Institute of Computing Technology of the Chinese Academy of Sciences. Her research interests include mobile/wireless networks, game theory, crowdsourcing.
\end{IEEEbiography}
\begin{IEEEbiography}[{\includegraphics[width=1in,height=1.25in,clip,keepaspectratio]{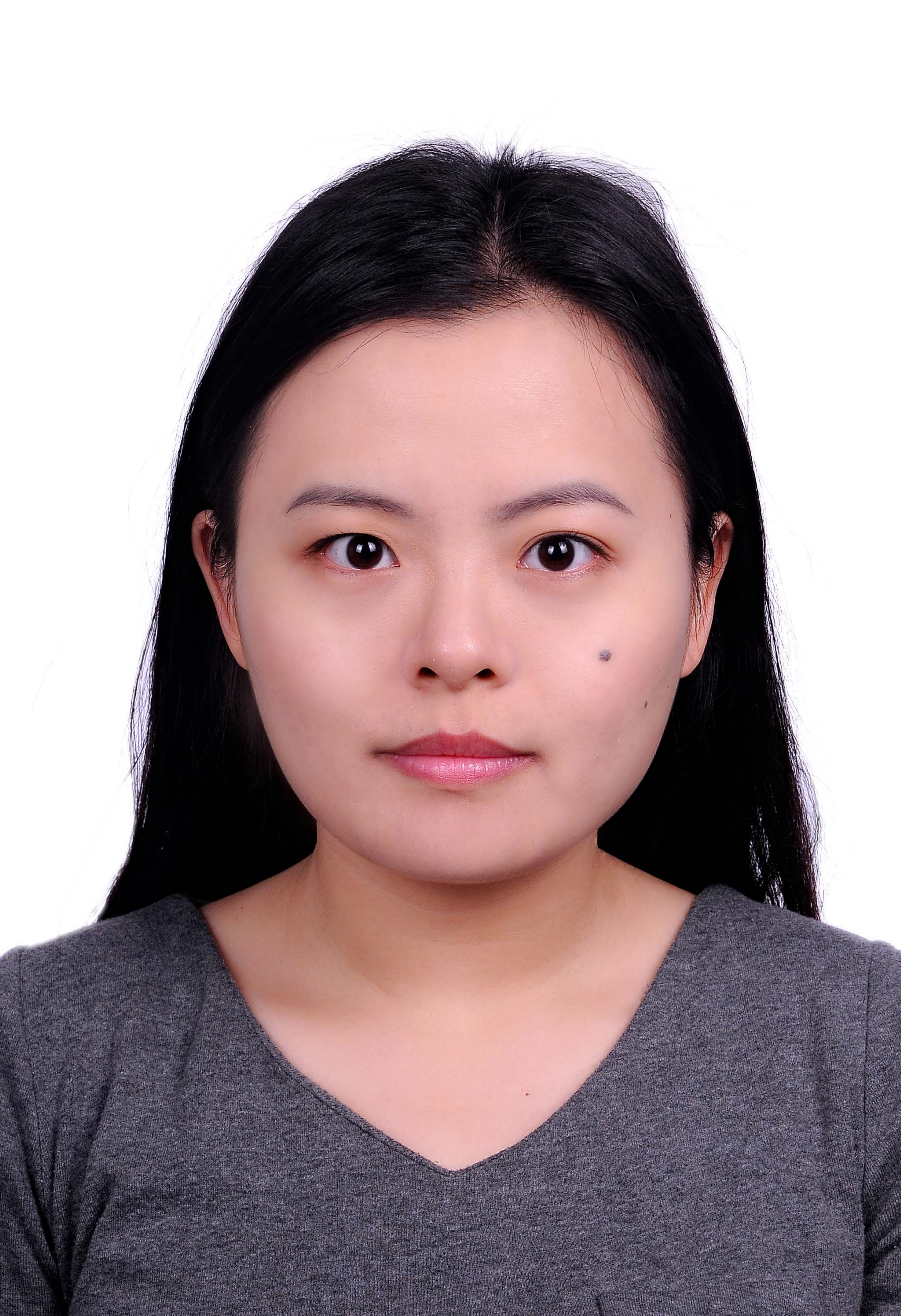}}]
{Qin Hu} received her Ph.D. degree in Computer Science from the George Washington University in 2019. She is currently an Assistant Professor in the department of Computer and Information Science, Indiana University - Purdue University Indianapolis. Her research interests include wireless and mobile security, crowdsourcing/crowdsensing and blockchain.
\end{IEEEbiography}

\begin{IEEEbiography}[{\includegraphics[width=1in,height=1.25in,clip,keepaspectratio]{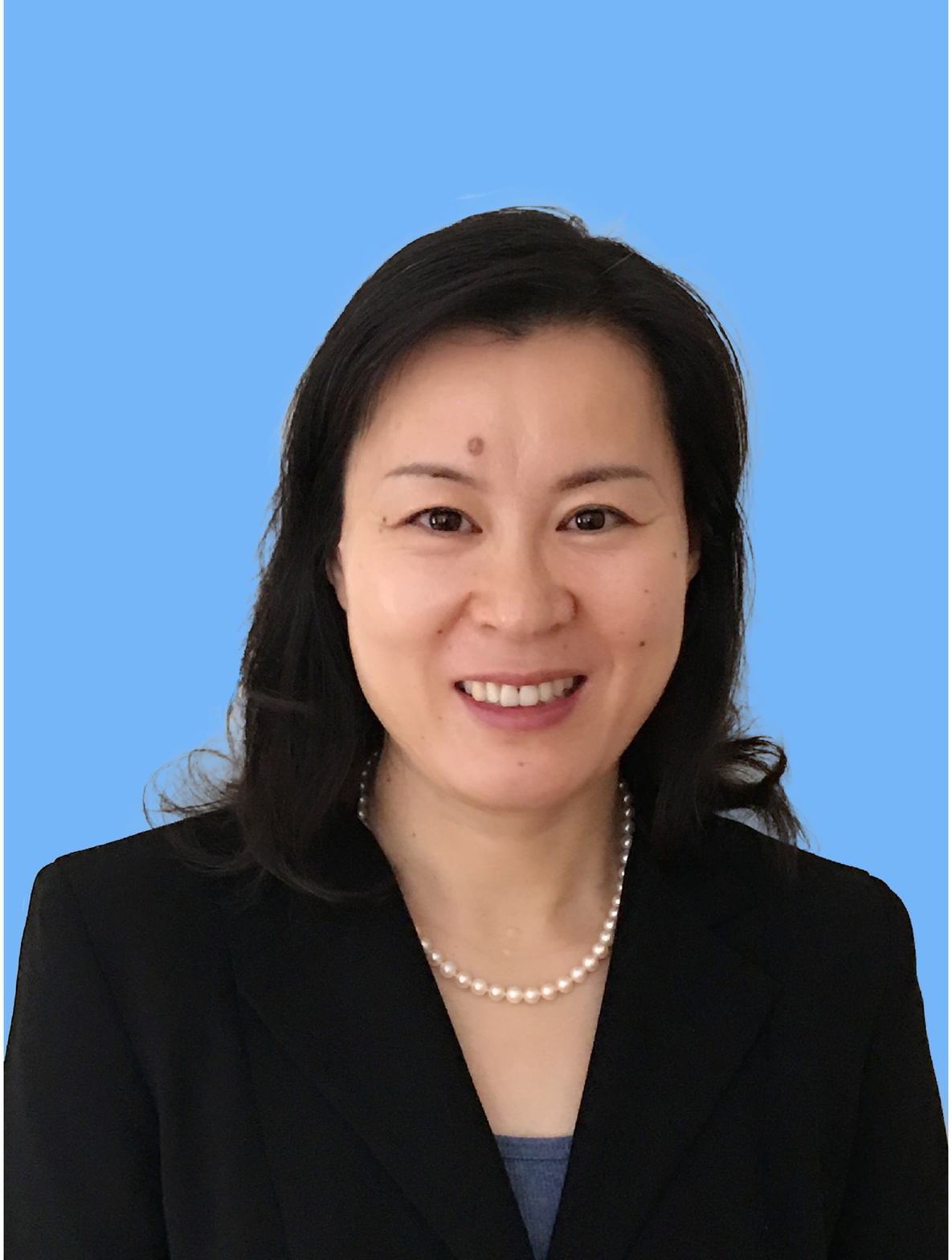}}]{Xiuzhen Cheng} received her M.S. and Ph.D. degrees in computer science from the University of Minnesota Twin Cities in 2000 and 2002, respectively. She is a professor in the School of Computer Science and Technology, Shandong University. Her current research interests focus on privacy-aware computing, wireless and mobile security, dynamic spectrum access, mobile handset networking systems (mobile health and safety), cognitive radio networks, and algorithm design and analysis.
 She has served on the Editorial Boards of several technical publications and the Technical Program Committees of various professional conferences/workshops. She has also chaired several international conferences. She worked as a program director for the U.S. National Science Foundation (NSF) from April to October 2006 (full time), and from April 2008 to May 2010 (part time). She published more than 170 peerreviewed papers.
\end{IEEEbiography}

\end{document}